\documentclass[]{aa}  

\usepackage{graphicx,siunitx}
\usepackage{txfonts}
\usepackage[normalem]{ulem}
\usepackage[]{hyperref}
\hypersetup{backref=true, pagebackref=true,hyperindex=true,breaklinks=true,colorlinks=true,urlcolor=blue,linkcolor=blue,citecolor=blue,pagecolor=red,bookmarks=true,bookmarksopen=true}

\begin{document}

\title{The SVS13-A Class I chemical complexity as revealed by S-bearing species. SOLIS XIII}


\author{C. Codella \inst{1,2}
\and 
E. Bianchi \inst{2,1} 
\and 
L. Podio \inst{1}
\and
S. Mercimek \inst{1,3}
\and
C. Ceccarelli \inst{2} 
\and
A. L\'opez-Sepulcre \inst{2,4}
\and
R. Bachiller \inst{5}
\and
P. Caselli \inst{6}
\and
N. Sakai \inst{7}
\and
R. Neri \inst{4}
\and
F. Fontani \inst{1}
\and
C. Favre \inst{2} 
\and
N. Balucani \inst{8,1,2}
\and
B. Lefloch \inst{2}
\and
S. Viti \inst{9,10}
\and
S. Yamamoto \inst{11,12}
}

\institute{
INAF, Osservatorio Astrofisico di Arcetri, Largo E. Fermi 5,
50125 Firenze, Italy
\and
Univ. Grenoble Alpes, CNRS, Institut de
Plan\'etologie et d'Astrophysique de Grenoble (IPAG), 38000 Grenoble, France
\and
Universit\`a degli Studi di Firenze, Dipartimento di Fisica e Astronomia, Via G. Sansone 1, 50019 Sesto Fiorentino, Italy
\and
Institut de Radioastronomie  Millim\'etrique, 300 rue de la Piscine, Domaine
Universitaire de Grenoble, 38406, Saint-Martin d'H\`eres, France
\and
Observatorio Astron\'omico Nacional (OAN-IGN), Alfonso XII 3,
28014 Madrid, Spain
\and
Max-Planck-Institut f\"ur extraterrestrische Physik (MPE), 
Giessenbachstrasse 1, 85748 Garching, Germany
\and
RIKEN Cluster for Pioneering Research, 2-1, Hirosawa, Wako-shi, 351-0198 Saitama, Japan
\and
Dipartimento di Chimica, Biologia e Biotecnologie, Via Elce di Sotto 8, 06123 Perugia, Italy
\and
Leiden Observatory, Leiden University, PO Box 9513, 2300 RA Leiden, The Netherlands
\and
Department of Physics and Astronomy, University College London, Gower Street, London, WC1E 6BT, UK
\and
Department of Physics, The University of Tokyo, 7-3-1, Hongo, Bunkyo-ku, Tokyo 113-
0033, Japan
\and
Research Center for the Early Universe, The University of Tokyo, 7-3-1, Hongo, Bunkyo-
ku, Tokyo 113-0033, Japan
}

\offprints{C. Codella, \email{codella@arcetri.inaf.it}}
\date{Received date; accepted date}

\authorrunning{Codella et al.}
\titlerunning{S-bearing molecules in SVS13-A}

\abstract{Recent results in astrochemistry revealed that some molecules 
such as interstellar complex organic species and deuterated species represent precious tools
to investigate star forming regions. Sulphuretted
species can also be used to follow the chemical evolution of the early stages of 
the Sun-like star forming process.}
{The goal is to obtain a census of S-bearing species using interferometric images, towards SVS13-A, a Class I object associated with a hot corino rich in interstellar complex organic molecules.}
{To this end, we used the NGC1333 SVS13-A data at3mm and 1.4mm obtained with the IRAM-NOEMA
interferometer in the framework of the Large Program SOLIS. The line emission of S-bearing species has been imaged and analysed using LTE and LVG approaches.} 
{We imaged the spatial distribution on $\leq$ 300 au scale of the line emission of  $^{32}$SO, $^{34}$SO, C$^{32}$S, C$^{34}$S, C$^{33}$S, OCS,  H$_2$C$^{32}$S, H$_2$C$^{34}$S, and NS. The low  excitation (9 K)
$^{32}$SO line is tracing (i) the
low-velocity SVS13-A outflow and (ii) the fast (up to 100 km s$^{-1}$ away from
the systemic velocity) collimated jet driven by the nearby SVS13-B Class 0 object.
Conversely, the rest of the lines are confined in the inner SVS13-A region, where
complex organics have been previously imaged.
More specifically the non-LTE LVG analysis of SO, SO$_2$, and
H$_2$CS indicates a hot corino origin (size in the 60-120 au range).
Temperatures between 50 K and 300 K, and volume densities larger than 
10$^{5}$ cm$^{-3}$ have been derived.
The abundances of the sulphuretted are in the following ranges:
0.3--6 $\times$ 10$^{-6}$ (CS), 
7 $\times$ 10$^{-9}$ -- 1 $\times$ 10$^{-7}$  (SO),
1--10 $\times$ 10$^{-7}$ (SO$_2$), a few 10$^{-10}$ 
(H$_2$CS and OCS), and 10$^{-10}$--10$^{-9}$ (NS).
The N(NS)/N(NS$^+$) ratio is larger than 10, 
supporting that 
the NS$^{+}$ ion is mainly formed in the extended envelope.}
{The [H$_2$CS]/[H$_2$CO] ratio, once measured at high-spatial resolutions,
increases with time (from Class 0 to Class II objects) by more than one order 
of magnitude (from $\leq$ 10$^{-2}$ to a few 10$^{-1}$).
This suggests 
that [S]/[O] changes along the Sun-like star forming process.
Finally, the estimate of the [S]/[H] budget in SVS13-A is
2\%-17\% of the Solar System value (1.8 $\times$ 10$^{-5}$),
being consistent with what was previously measured towards 
Class 0 objects (1\%-8\%). This supports
that the enrichment of the sulphuretted species with respect to dark clouds
keeps constant from the Class 0 to the Class I stages of low-mass star 
formation. The present findings stress the importance of
investigating the chemistry of star forming regions using
large observational surveys as well as sampling regions on a Solar System scale.}


\maketitle

\section{Introduction}

Molecular complexity builds up at each step of the process leading to Sun-like star formation.
Searches for exoplanets have shown a large degree of diversity in the planetary systems\footnote{https://exoplanets.nasa.gov/}, and it is unclear how common a System like our own is.
In this context, the role of the pre-solar chemistry in the formation of the Solar System 
bodies is far from being understood. A breakthrough question 
\citep[see e.g.][and references therein]{Ceccarelli2007,Herbst2009,Caselli2012,Jorgensen2020} is to understand whether planetary systems 
inherited at least part of the chemical composition of the earliest 
stages of the Sun-like star forming process, such as prestellar cloud, Class 0 ($\leq$ 10$^5$ yr),
Class I ($>$ 10$^5$ yr), and Class II ($>$ 10$^6$ yr) objects.
To progress, it is of paramount importance:
(i) to observe objects representing all the evolutionary
stages of the star forming process sampling regions 
on a Solar System scale, and (ii) to 
combine high-sensitivity spectral surveys collecting a large number of species and emitting lines to constrain their abundances.  
  
In the last years, ALMA revolutionised our comprehension of planet formation, delivering images of rings and gaps 
\citep[see e.g.,][]{Sheehan2017,Fedele2018,Andrews2018} in the dust distribution around objects with an age less than 1 Myr. This
clearly supports that planet formation occurs earlier
than what was postulated before, more specifically
already in the Class I phase. This in turn stresses the importance of
investigating Class I objects, a sort of bridge between the prototellar phase and the protoplanetary disks around Class II young objects.
Several projects focused on the chemical complexity of 
Class I objects have been recently performed, mainly focusing
the attention on interstellar small or complex (with at least 6 atom) organic molecules
and/or deuteration using both single-dishes and interferometer
\citep[see e.g.,][]{Oberg2014,Bianchi2017a,Bergner2017,Bergner2019,Bianchi2019a, Bianchi2019b,Bianchi2020,Legal2020,Yang2021}.

In the dense gas typical of star forming regions,
[S]/[H] is far to the value measured in our Solar System
\citep[1.8 $\times$ 10$^{-5}$][]{Anders1989}, given
sulphur is depleted by several orders of magnitude \citep[e.g.][]{Tieftrunk1994,Wakelam2004,Phuong2018,
Laas2019,vantHoff2020}.
On grains, sulphur is expected to be very refractory, likely in the form
of FeS \citep{Kama2019} or S$_{\rm 8}$ \citep{Shing2020}, while
it is still debated which is the main reservoir on dust mantles.
For years, H$_2$S and possibly OCS have been postulated to be the solution, but so far they have never been directly detected in interstellar ices \citep{Boogert2015}, so that it seems that other, not yet identified, frozen species contain the majority of sulphur.

The abundance of gaseous S-bearing species drastically increases in the regions where the species frozen out onto the dust mantles are injected into the gas-phase either due to thermal
evaporation in the heated central zone of protostars  (e.g. in the Class 0/I hot corinos) or because of gas-grain sputtering
and grain-grain shattering in shocked regions
\citep[e.g.][]{Charnley1997,Bachiller2001, Wakelam2004,Codella2005,Sakai2014a,Sakai2014b,Podio2014,Imai2016,Holdship2016,Taquet2020,Feng2020}.
In both cases, S-bearing species have proved to be extremely useful in the reconstruction of both the chemical history and dynamics of the studied objects.
In addition, S-bearing species have been recently imaged in relatively older protoplanetary disks using CS, SO, H$_2$S, and H$_2$CS line emission \citep[e.g.][]{Teague2018,Booth2018,Phuong2018,LeGal2019,Loomis2020,Codella2020,Garufi2020,Garufi2021,Podio2020-iras,Podio2020-dgtau,Oya2020}. 
All these studies (i) have shown that S-bearing species are a powerful tool to follow the evolution of the chemistry until the latest stages of star formation, and (ii) call for more studies of S-bearing species towards the Class I protostars, which, as mentioned above, represent a crucial transition from the youngest Class 0 protostars and the more evolved Class II protoplanetary disk sources.

In the present paper, we report a survey of the S-bearing species in the Class I source prototype SVS13-A observed in the framework of the IRAM-NOEMA SOLIS (Seeds Of Life In Space)\footnote{\url{https://solis.osug.fr/}} \citep{ceccarelli2017}.
The article is organised as follows.
In Sect. \ref{sec:svs13background}, we report the description of target source, SVS13-A.
We describe the observations in Sect. \ref{sec:obs} and the results in Sect. \ref{sec:results}.
In Sect. \ref{sec-phys-parameters}, we carry out a non-LTE analysis of the obtained data and derive the physical and chemical parameters of the detected S-bearing in the molecules emitting gas.
We discuss the implications of our findings in Sect. \ref{sec:discussion} and summarise the conclusions of this work in Sect. \ref{sec:conclusions}.

\section{The SVS13-A Class I prototypical source}\label{sec:svs13background}

The SVS13-A object is located in the  NGC1333 cluster, in the Perseus region, at a distance of 299 $\pm$ 14 pc  \citep{Zucker2018}.
SVS13-A is in turn a 0$\farcs$3 binary source (VLA4A, VLA4B; \citealt{Anglada2000, Tobin2016, Tobin2018}), so far not disentangled using mm-wavelengths observations (e.g. \citealt{Lefevre2017,Maury2019}),
located close ($\sim$ 4$\arcsec$) is a third object called VLA3.
The three objects are surrounded by a large 
molecular envelope \citep{Lefloch1998a}.
The bolometric luminosity of SVS13-A is $\sim$ 50 $L_{\rm bol}$,
and it is considered as one of the archetypical Class I sources (at least 10$^5$ yr, e.g. \citealt{Chini1997}), having been observed in the last decades at different spectral windows
(see e.g. \citealt{Chini1997, Bachiller1998, Looney2000, Chen2009, Tobin2016, Maury2019}, and references therein).
SVS13-A is driving an extended molecular outflow \citep{Lefloch1998a, Codella1999, Sperling2020, Dionatos2020}, associated with the Herbig-Haro (HH) chain 7--11 \citep{Reipurth1993} as well as younger flows moving towards 
Southern-Eastern directions \citep{Lefevre2017}.

A hot corino has been revealed towards SVS13-A using deuterated water 
observed with the IRAM 30-m by \citet{Codella2016} and then imaged 
by \citet{Desimone2017} using the IRAM PdBI and HCOCH$_2$OH (glycolaldehyde) 
emission lines, finding a 90 au diameter. The first iCOMs census
of interstellar Complex Organic Species (iCOMs; organic molecules
with at least 6 atoms) has been reported by \citet{Bianchi2019a} thanks to the ASAI (Astrochemical Surveys At IRAM) Large Program \citep{Lefloch2017} 
unbiased survey in the IRAM spectral windows. In this context, 
\citet{Bianchi2017a, Bianchi2019b} analysed fractionation and deuteration of a large number of species. Several iCOMs have been also 
detected by \citet{Belloche2020} using high-resolution PdBI observations.  
Unfortunatly, even the interferometric campaigns did not
allow the observers to clearly identify which component of the binary system
is associated with a rich chemistry, noting only that the emission peak is offset ($\sim$ 1$\farcs$5 to West) from VLA4A \citep[see also][]{Lefevre2017}.

\section{Observations}\label{sec:obs}

The SVS13-A region was observed at 1.4mm and 3mm in two different setups \citep[hereafter labelled 5 and 6: see also:][]{ceccarelli2017} with the IRAM NOEMA interferometer.
The phase
center of the obtained images is $\alpha_{2000}$ = 03$^h$29$^m$03$^s$.73, $\delta_{2000}$ = +31$^\circ$16$'$03$''$.8.

Setup 6, at 3mm, was observed in A configuration 
(1 track) using 9 antennas in March 2018. 
The frequency ranges are 80.2--88.3 GHz and
95.7--103.9 GHz.
The shortest and longest projected baselines are 64 m and 760 m, respectively. 
The field of view is about 60$\arcsec$, while the Largest Angular Scale (LAS) is $\simeq$ 8$\arcsec$. Line images were produced by subtracting continuum image, 
using natural weighting, and restored with 
a clean beam, for continuum, of 1$\farcs$8 $\times$ 1$\farcs$1 (PA $\simeq$ 41$^\circ$).
Note that the scope of the present project is to focus on the hot corino and not the extended emission. 
The Polyfix correlator was used, with a total
spectral band of about 8 GHz, and a spectral resolution of 2 MHz ($\simeq$ 7 km s$^{-1}$). 
Setup 5, at 1.4mm, was observed in A and C configurations  
(3 tracks) using 8 antennas in December 2016.
The frequency range is 204.0--207.6 GHz.
The shortest and longest projected baselines are 24 m and 760 m, respectively, for a field of view of 24$\arcsec$ and LAS $\sim$ 9 $\arcsec$.
The clean beam is, for the continuum at 1.4 mm,
0$\farcs$6 $\times$ 0$\farcs$6 (PA $\simeq$  --46$^\circ$).
The WideX backend has been used providing a bandwidth of $\sim$ 3.6 GHz with a spectral resolution of 
$\sim$ 2 MHz (2.8--2.9 km s$^{-1}$). 
In addition, 320 MHz wide narrowband backends
providing a spectral resolution of $\sim$
0.9 km s$^{-1}$ have been used.

For both setups, 
calibration was carried out following standard
procedures, using GILDAS-CLIC\footnote{http://www.iram.fr/IRAMFR/GILDAS}.
The bandpass was fixed on 3C84, 
while the absolute flux was calibrated on 
LkH$\alpha$101, MWC249, and 0333+321.
The final uncertainty on the absolute flux scale 
is $\leqslant$ 10$\%$ (3mm) and $\leqslant$ 15$\%$ (1.4mm). 
The phase rms
was $\leqslant$ 50$^\circ$, and the typical precipitable water vapor (pwv) 
was $\sim$ 5-15 mm. Finally, 
the system temperature was $\sim$ 50-200 K. 
The final rms noise in the broadband cubes 
is $\sim$ 20--50 mJy beam$^{-1}$ (3mm),
and $\sim$ 700--1000 mJy beam$^{-1}$ (1.4mm). 

Figures A.1-A.4 compares the NOEMA-SOLIS spectra with those derived at the
same frequency with the IRAM 30-m \citet{Lefloch2018},  smoothed to the 2 MHz
spectral resolution of the SOLIS data). Red labels indicate the S-bearing species analysed in this paper (see Table 1).
As expected, the single-dish observations are detecting the photons emitted
by the multiple components inside the IRAM 30-m HPBW (25$\arcsec$--30$\arcsec$ at
3mm, and 12$\arcsec$ at 1.4mm), such as the cold extended envelope or
the large-scale SVS13-A outflow. On the other hand, NOEMA-SOLIS, filtering out the emission at scales larger than $\sim$ 8$\arcsec$ is well suited for the
analysis of the inner 100 au protostellar region as long as the filtering does not affect the line profile which traces the hot corino.

\section{Results}\label{sec:results}

\subsection{Continuum images}

Figure 1--Left shows the the SVS13-A region as observed in dust continuum emission at 3mm. Several protostars are detected: SVS13-A (without disentangling the binary
components VLA4A and 4B at the present 1$\farcs$3 angular resolution), VLA3, SVS13-B, and, out of the primary beam, also SVS13-C. Figure 1-Right shows the 1.4mm map of the 
inner 30$\arcsec$ (the primary beam is, in this case, 24$\arcsec$): while there is a tentative 5$\sigma$ signature of VLA3, SVS13-A and SVS13-B are clearly
detected. Still, at an angular resolution of 0$\farcs$6,
the SVS13-A binary system (separated by 0$\farcs$3) is not disentangled.

The J2000 coordinates of the protostars are
in agreement with the previous continuum imaging
in both cm- and mm-wavelengths
(see e.g. \citealt{Looney2000, Anglada2000, Tobin2018, Maury2019}), namely:
SVS13-A: 03$^h$29$^m$03$^s$.757, +31$^\circ$16$'$03$\farcs$74; VLA3: 03$^h$29$^m$03$^s$.386, +31$^\circ$16$'$01$\farcs$56;
SVS13-B: 03$^h$29$^m$03$^s$.064, +31$^\circ$15$'$51$\farcs$50; SVS13-C: 03$^h$29$^m$01$^s$.947, +31$^\circ$15$'$37$\farcs$71.
Finally, the peak fluxes, for the sources imaged inside the primary beams, are: SVS13-A: 
26.6$\pm$0.1 mJy beam$^{-1}$ (90 GHz), and 94$\pm$1
mJy beam$^{-1}$ (205 GHz); VLA3:
2.2$\pm$0.1 mJy beam$^{-1}$ 
(90 GHz), and 5$\pm$1 mJy beam$^{-1}$ (205 GHz);
SVS13-B: 19.2$\pm$0.1 mJy beam$^{-1}$ (90 GHz).

\begin{figure*}
\centerline{\includegraphics[angle=0,width=17.5cm]{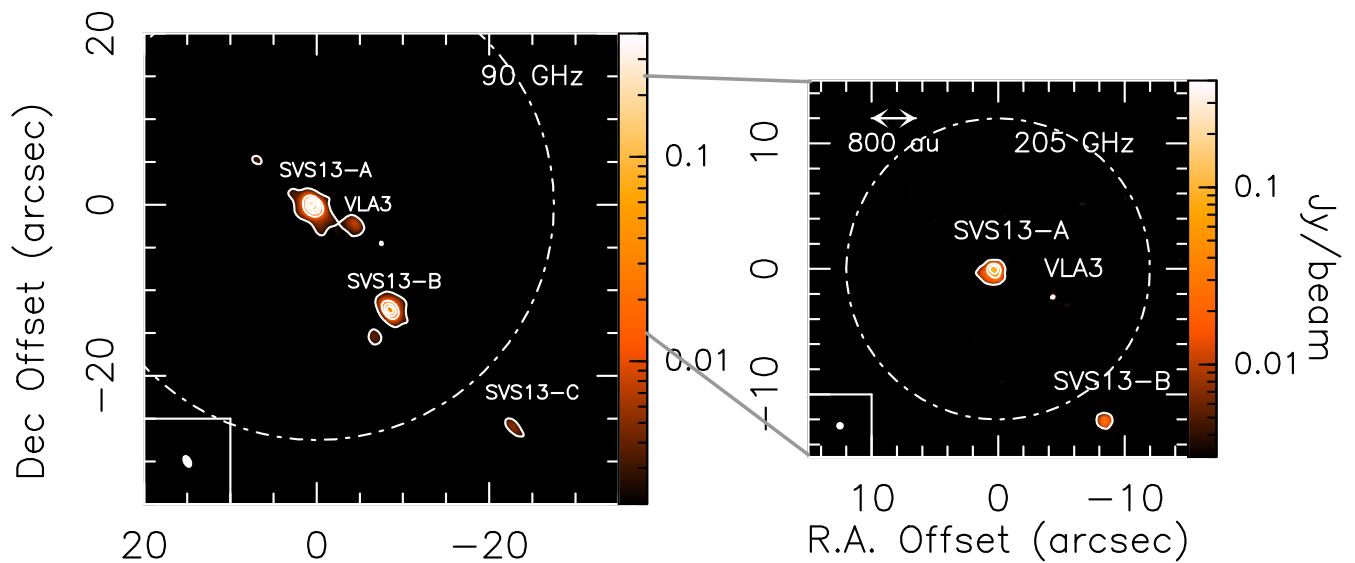}}
\caption{{\it Left Panel}: The SVS13-A region as observed in dust continuum emission at 3mm (see Sect. 2) using IRAM NOEMA.
Angular offset are with respect to the phase center
(see Sect. 3).
First contour and steps are, respectively, 5$\sigma$ (1$\sigma$ = 100 $\mu$Jy) and 50$\sigma$.
The filled ellipse shows the synthesised beam (HPBW): 1$\farcs$6 $\times$ 1$\farcs$1 (PA= 41$^\circ$).
The SVS13-B, SVS13-C, and VLA3 protostars are also labelled. White dashed circle indicate the primary beam:
$\sim$ 55$\arcsec$.
{\it Right Panel}: Zoom-in of the inner SVS13-A region as observed in dust continuum at 1.3mm. 
First contour and steps are, respectively, 5$\sigma$ (1$\sigma$ = 1 mJy) and 30$\sigma$.
Symbols are as in the Left panel.
The synthesised beam (HPBW) is 0$\farcs$6 $\times$ 0$\farcs$6 (PA= --46$^\circ$), while the primary beam is 24$\arcsec$.}
\end{figure*}

\subsection{Line images and spectra}

We imaged, at both 1.4 mm and 3mm, 
a large (32) number of lines of
S-bearing species, listed in Table 1.
Namely, 4 $^{32}$SO (hereafter SO) lines ($E_{\rm u}$ in the 9--39 K range),
2 $^{34}$SO lines (9--19 K), 7 SO$_2$ lines (37--549 K),
2 $^{34}$SO$_2$ lines (55--70 K), 1 C$^{32}$S (hereafter CS) line (7 K), 1 C$^{34}$S line (6 K), 1 C$^{33}$S line (7 K), 3 OCS lines (16--89 K), 9
H$_2$C$^{32}$S (hereafter H$_2$CS) lines (including two pairs blended at the present spectral resolution; 10--244 K), 1 H$_2$C$^{34}$S line
(48 K), and 1 NS line (27 K). In the following, 
for sake of clarity, the
results of each species will be reported separately.
The overall picture will be discussed in Sect. 6.

\subsubsection{SO}

Figure 2-Left shows the spatial distribution of the
low-$E_{\rm u}$ (9 K) SO(2$_{\rm 3}$--1$_{\rm 2}$) emission
observed at 86 GHz. The emission peaks are towards SVS13-A 
and SVS13-B (3mm continuum drawn in black) and are spatially unresolved. In addition, a contribution from 
the extended envelope is suggested.  
Figure 3 reports the SVS13-A spectrum.
The used low-spectral resolution (7 km s$^{-1}$) 
prevents us from a proper kinematical analysis. 
However, the peak velocity is centred at the systemic velocity of +8.6 km s$^{-1}$ \citep{Chen2009}.
By imaging the SO(2$_{\rm 3}$--1$_{\rm 2}$) per
velocity range, Fig. 2-Left shows
some blue-shifted emission flowing towards South-East,
i.e. tracing signatures of the well-known extended 
($\gg$ 8$\arcsec$) outflow associated with HH7-11
(see e.g. \citealt{Lefloch1998a, Codella1999, Lefevre2017}).
In addition, a well collimated (and more compact) bipolar outflow
driven by SVS13-B is observed. 
Very high-velocity emission, up to +20 km s$^{-1}$ and down to --94 km s$^{-1}$, is detected.
This outflow, located along the NW-SE
direction, had been discovered by
\citep{Bachiller1998} using SiO, and recently 
imaged in the context of the CALYPSO IRAM Large Program \citet{Maury2019} by 
\citet{Podio2021} in the SiO(5--4), CO(2--1), and SO(5$_{\rm 6}$-4$_{\rm 5}$) lines.
The outflow collimation is consistent with SVS13-B being in an earliest  evolutionary stage (Class 0) with respect to SVS13-A (Class I) driving a more extended, and less collimated flow.

Moving towards slightly higher excitation lines (up to 39 K), the SO lines
observed at 3mm with $\sim$ 1$\farcs$5 (450 au) angular resolution appear spatially unresolved and 
centred on SVS13-A. SVS13-B is detected only through the 2$_{\rm 2}$--1$_{\rm 1}$ and 2$_{\rm 3}$--1$_{\rm 2}$ lines. On the other hand, 
the images of the SO lines at 1.4mm
(synthesised beam $\sim$ 180 au) reveal a structure with a size $\sim$ 300 au, plausibly associated with the  molecular envelope. An elongation (due to low-velocity blue-shifted emission) towards the blue-shifted outflow
SE direction is clearly visible. The SO(4$_{\rm 5}$--3$_{\rm 4}$) line
($E_{\rm u}$ = 39 K) profile, observed with a spectral resolution 
of 2.8 km s$^{-1}$, has a FWHM line width of 9.2 km s$^{-1}$.

Finally, the ratio between the integrated fluxes 
as derived towards SVS13-A
of the SO and $^{34}$SO 2$_{\rm 3}$--1$_{\rm 2}$ or
2$_{\rm 2}$--1$_{\rm 1}$ lines is $\sim$10--12, which, assuming
optically thin $^{34}$SO emission and $^{32}$S/$^{34}$S = 22 \citep{WilsonRood}, leads to an SO opacity $\sim$ 1.

\begin{figure*}
\centerline{\includegraphics[angle=0,width=17.5cm]{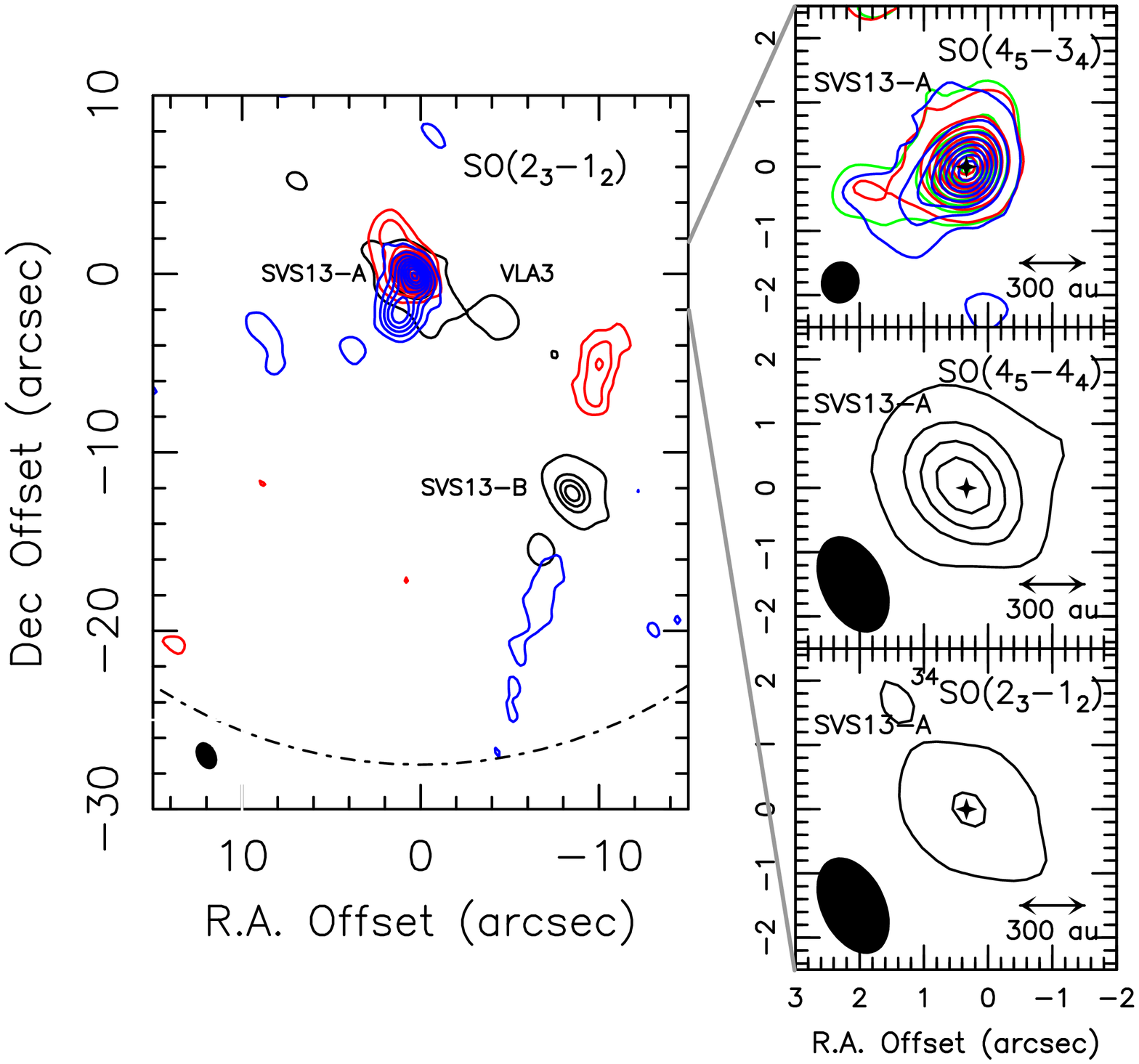}}
\caption{{\it Left Panel}: Spatial distribution of the
SVS13-A region obtained using the SO(2$_{\rm 3}$--1$_{\rm 2}$) red- and blue-shifted emission overlaid with the 90 GHz continuum map (see Fig. 1). 
Red-shifted emission maps have been derived by integrating the SO emission from the systemic velocity
(+8.6 km s$^{-1}$, \citet{Chen2009}) up to
+20 km s$^{-1}$. Blue-shifted image has been obtained 
by integrating down to --94 km s$^{-1}$ in absolute values).
Angular offset are with respect to the phase center 
(see Sect. 3).
First contour and steps are, respectively, 3$\sigma$,
and 4$\sigma$, respectively.
The 1$\sigma$ value is 5 mJy km s$^{-1}$ and 14 mJy km s$^{-1}$ for the red- and blue-shifted emission, respectively.
The filled ellipse shows the synthesised beam (HPBW): 1$\farcs$6 $\times$ 1$\farcs$1 (PA= 41$^\circ$).
The SVS13-B protostar is also labelled. White dashed circle indicate the primary beam: $\sim$ 55$\arcsec$.
{\it Upper-Right Panel}: Zoom-in of the inner SVS13-A region as observed using the SO(4$_{\rm 5}$--3$_{\rm 4}$) emission.
Green contours are for the emission in the 3 km s$^{-1}$ around the systemic
velocity.
Red-shifted emission maps has been derived by integrating the SO emission from the systemic velocity up to
+20 km s$^{-1}$. Blue-shifted image has been obtained 
by integrating down to --5 km s$^{-1}$.
First contour and steps are, respectively, 3$\sigma$, and 10$\sigma$.
The 1$\sigma$ value is 10 mJy km s$^{-1}$ (systemic velocity), 13 mJy km s$^{-1}$ (red), and 11 mJy km s$^{-1}$ (blue).
Symbols are as in the Left panel.
The black cross indicates the positions of the 205 GHz continuum
peak (see Fig. 1).
The synthesised beam (HPBW) is 0$\farcs$65 $\times$ 0$\farcs$58 (PA= --46$^\circ$).
{\it Middle-Right Panel}: Zoom-in of the inner SVS13-A region as observed using the SO(4$_{\rm 5}$--4$_{\rm 4}$) emission. The whole velocity emitting region has been used (see Fig. 3). First contour and steps are, respectively, 3$\sigma$ (9 mJy km s$^{-1}$), and 10$\sigma$. Symbols are as in the Left panel. The synthesised beam (HPBW) is: 1$\farcs$6 $\times$ 1$\farcs$1 (PA= 41$^\circ$).
{\it Bottom-Right Panel}: Zoom-in of the inner SVS13-A region as observed using the $^{34}$SO(2$_{\rm 3}$--1$_{\rm 2}$) emission. First contour and steps are, respectively, 3$\sigma$ (12 mJy km s$^{-1}$), and 10$\sigma$. Symbols 
are as in the Left panel. The synthesised beam is as in the Middle-Right panel.}
\end{figure*}

\begin{figure*}
\centerline{\includegraphics[angle=0,width=15cm]{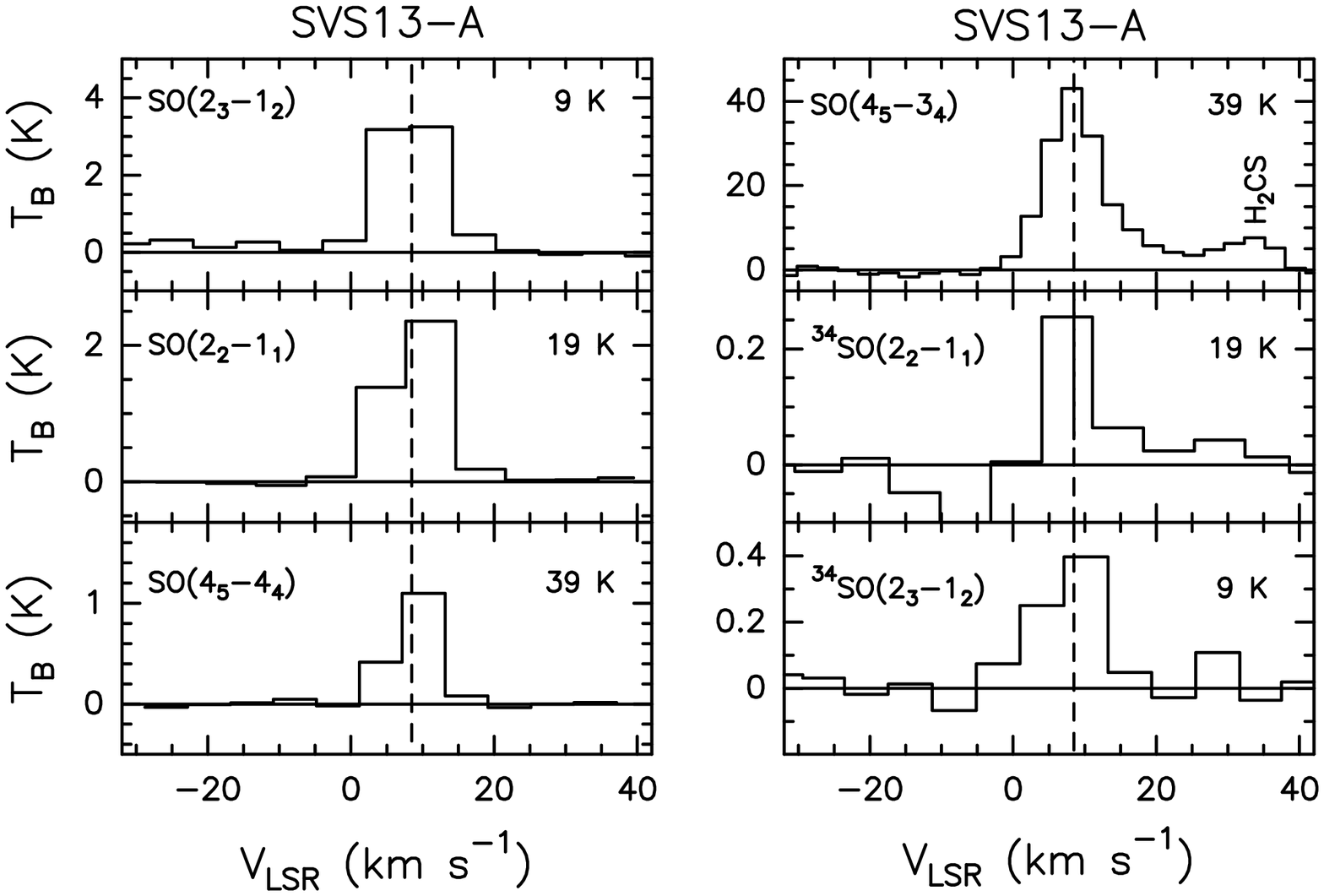}}
\caption{Observed SO, and $^{34}$SO spectra 
(in T$_{\rm B}$ scale, see Table 1) extracted at the emission peak: $\alpha_{2000}$ = 03$^h$29$^m$03$^s$.75,  $\delta_{2000}$ = +31$^\circ$16$'$03$''$.8. 
Transitions and corresponding upper level energies are reported.
The vertical dashed line stands for the ambient LSR velocity (+8.6 km s$^{-1}$, \citet{Chen2009}). 
Note that the p-H$_2$CS(6$_{\rm 2,4}$--5$_{\rm 2,3}$) appears close to the SO(4$_{\rm 5}$--3$_{\rm 4}$) profile (see Tab. 1).}
\end{figure*}

\subsubsection{CS}

We imaged the $J$ = 2--1 line emission of the CS, 
C$^{34}$S, and C$^{33}$S isotoplogues, emitting 
in the 96.4--98.0 GHz spectral range (Table 1).
Figure 4 shows how the main isotopologue is tracing the molecular
envelope around SVS13-A. The rarer isotopologues are 
emitting towards the CS emission peak, having a spatially unresolved
structure ($\leq$ 450 au). The spectra corresponding to 
the CS peak are drawn in Fig. 5. 
The line ratios observed towards SVS13-A are: CS/C$^{34}$S $\simeq$ 3 
and CS/C$^{33}$S $\simeq$ 11. In turn, assuming
the C$^{33}$S emission to be optically thin and $^{32}$S/$^{33}$S = 138 \citep{WilsonRood}, this implies optically thick ($\tau$ $\sim$ 14) CS emission.
The C$^{34}$S emission results to be moderately thick ($\tau_{\rm C^{34}S}$ $\simeq$ 2).

\begin{figure}
\centerline{\includegraphics[angle=0,width=7cm]{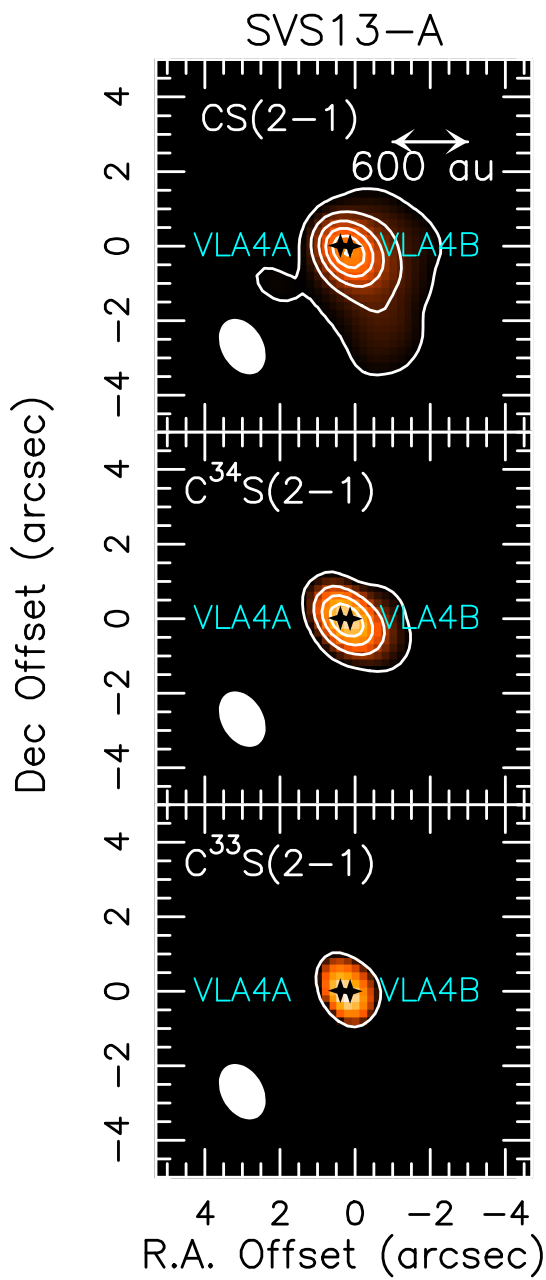}}
\caption{Spatial distribution of the
SVS13-A region obtained using the $J$ = 2--1 CS (Upper panel), C$^{34}$S
(Middle), 
and C$^{33}$S (Bottom) line emission (contours and colour). 
The whole velocity emitting region has been used 
(down to -- 5 km s$^{-1}$ and up to
+25 km s$^{-1}$, depending on the line, see Fig. 5). Angular offset are respect to the phase center 
(see Sect. 3).
First contour and steps are, respectively, 5$\sigma$,
and 10$\sigma$, respectively.
The 1$\sigma$ value is 10 mJy km s$^{-1}$ (CS), and 
4 mJy km s$^{-1}$ (C$^{34}$S, C$^{33}$S). 
The filled ellipse shows the synthesised beam (HPBW): 1$\farcs$6 $\times$ 1$\farcs$1 (PA= 41$^\circ$).
The black crosses indicate the positions of the VLA4A and VLA4B
sources as imaged using the VLA array by \citet{Tobin2018}.}
\end{figure}

\begin{figure}
\centerline{\includegraphics[angle=0,width=8cm]{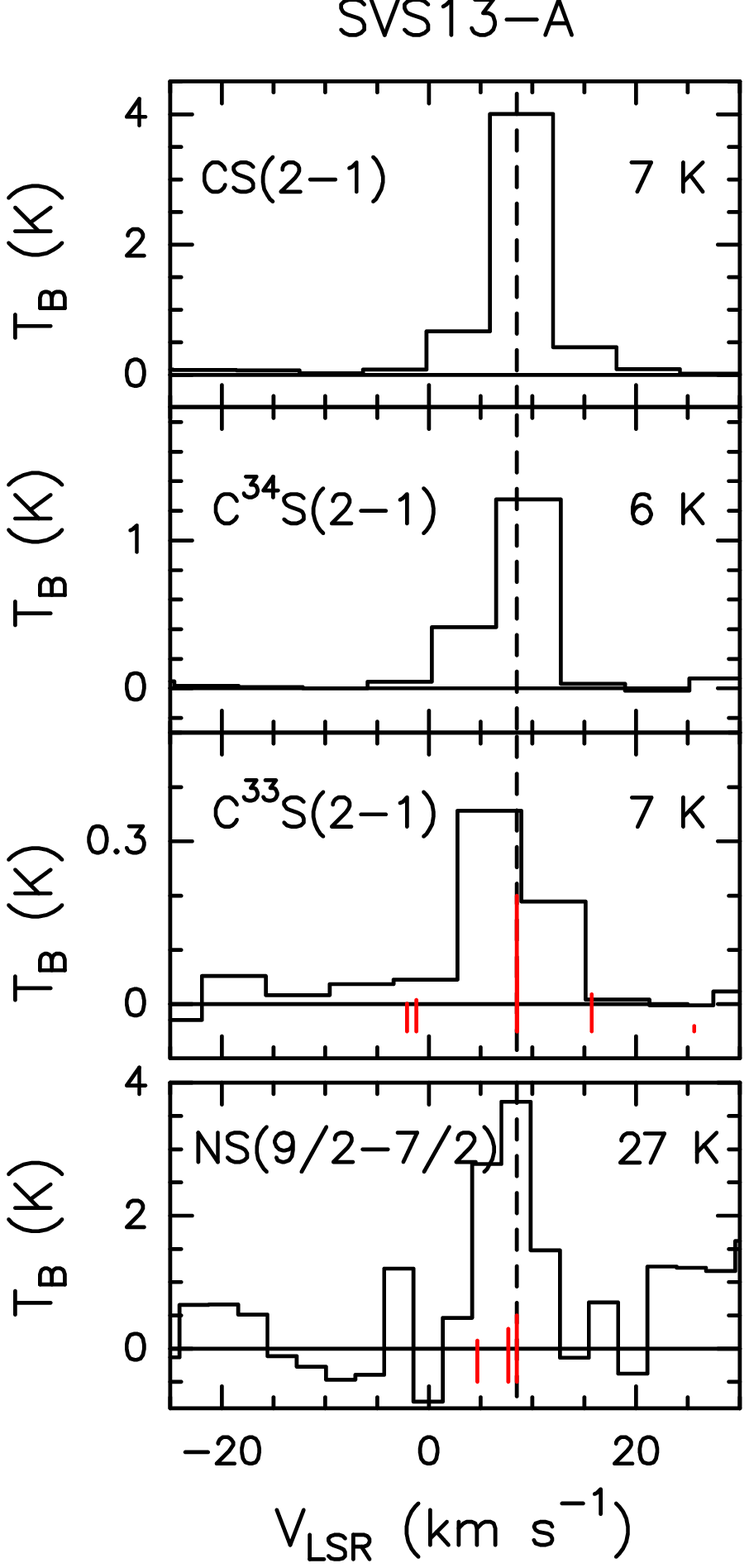}}
\caption{Observed CS, C$^{34}$S, C$^{33}$S, and NS spectra 
(in T$_{\rm B}$ scale, see Table 1) extracted at the emission peak: $\alpha_{2000}$ = 03$^h$29$^m$03$^s$.75,  $\delta_{2000}$ = +31$^\circ$16$'$03$''$.8. 
Transitions and corresponding upper level energies are reported.
The vertical dashed line stands for the ambient LSR velocity (+8.6 km s$^{-1}$, \citet{Chen2009}). 
The  C$^{33}$S(2--1) line consists of 6 hyperfine
components with $S\mu^2$ in the 0.5--12.2 D$^2$ range \citep{Bogay1981,Lovas2004,Muller2005} spread on a 9 MHz frequency interval. The vertical red lines (in scales according to their $S\mu^2$)
indicated their relative offset in velocity scale
with respect to the brightest $F$=5/2--3/2 line.
The NS 9/2--7/2 $\Omega$=1/2 line consists of 3 hyperfine
components \citep{Lee1995-NS} associated
with $S\mu^2$ in the 10.9--17.4 D$^{2}$ range 
spread on a 0.6 MHz frequency interval.
The vertical red lines (in scales according to their $S\mu^2$) 
indicated their relative offset in velocity scale
with respect to the brightest $F$=11/2--9/2 line (see Table 1).
}
\end{figure}

\subsubsection{SO$_2$}

A large number of SO$_2$ (and $^{34}$SO$_2$) lines 
has been detected at both 3mm and 1.4mm,
covering a very large range of the upper level energies,
from 37 K to 549 K. Figure 6 reports a selection of 
the spatial distributions of the  SO$_2$ lines, peaking towards SVS13-A.
All the spectra are reported in Fig. 7: when observed with a spectral
resolution of 2.9 km s$^{-1}$, the profiles peak close to the systemic
velocity (+8.6 km s$^{-1}$, \citealt{Chen2009}), and are $\sim$ 7--8 km s$^{-1}$ broad.

Contrarily to SO and CS, the SO$_2$ emission is spatially unresolved 
both at 3mm and 1.4mm, indicating an emitting size less than 180 au.
In other words, there is neither signature of outflows (as for SO) nor
of the envelope revealed by SO and CS. The SO$_2$ and
$^{34}$SO$_2$ emission
is tracing the inner protostellar region. By fitting the position 
of the brightest SO$_2$ line (204.2 GHz) 
using a Gaussian fit in the $uv$ domain\footnote{In the $uv$ domain, the error on centroid positions is the function of the channel signal-to-noise ratio and atmospheric seeing, and is typically much smaller than the beam size.}, we found 
$\alpha_{2000}$ = 03$^h$29$^m$03$^s$.755,  $\delta_{2000}$ = +31$^\circ$16$'$03$''$.782. The $uv$ fit error is 3 mas.
This position lies between
the coordinates of the binary components VLA4A and VLA4B 
(see Fig. 6-Right panels), more specifically 
at $\simeq$ 0$\farcs$13 (39 au) from both protostars.
This shift perfectly agrees with what was found using 
iCOMs emission as imaged with an 
angular resolution of 0$\farcs$4--1$\farcs$3 (at 1.3--3mm) in the 
context of the CALYPSO project \citep{Desimone2017,Belloche2020}.

Finally, since we have not detected common SO$_2$ and $^{34}$SO$_2$ lines, the estimate on the SO$_2$ opacity will be discussed in Sect. 5 in light
of the Large Velocity Gradient (LVG) analysis.


\begin{figure*}
\centerline{\includegraphics[angle=0,width=18cm]{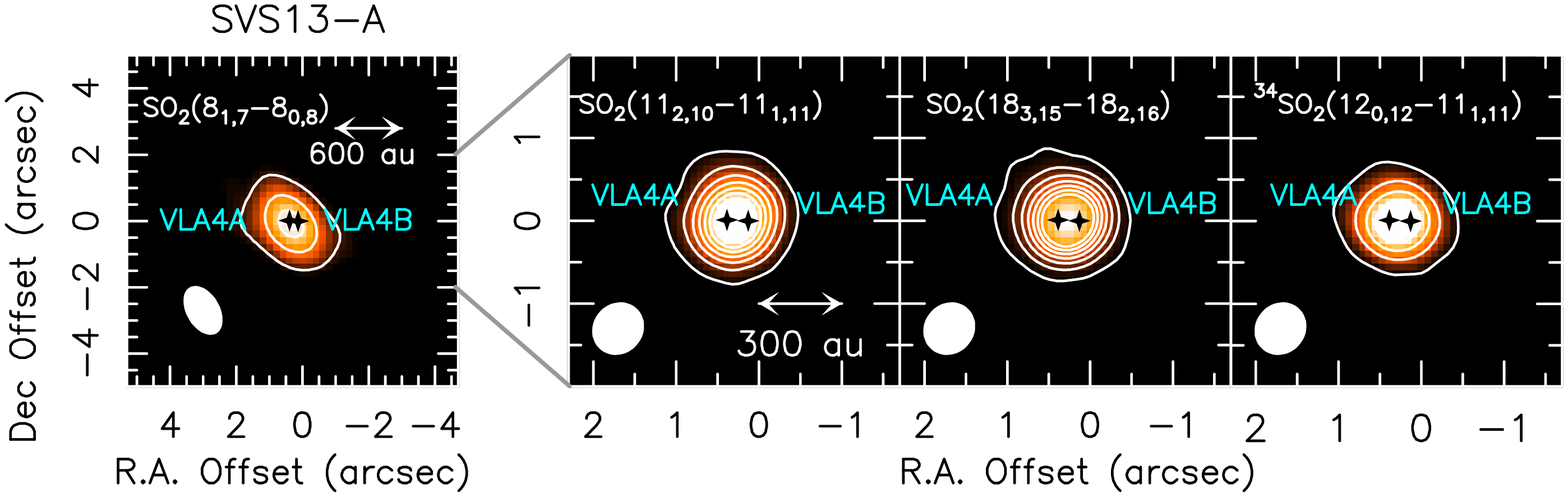}}
\caption{Spatial distribution (contours and colour) of the
SVS13-A region obtained using different SO$_2$, and
$^{34}$SO$_2$ line emission (see Table 1). 
Angular offset are respect to the phase center 
(see Sect. 3).
The whole velocity emitting region has been used (from 0 km s$^{-1}$
to +20 km s$^{-1}$, see Fig. 7). The SO$_2$(11$_{\rm 2,10}$--11$_{\rm 1,11}$) profile 
has been deblended by the CH$_3$OCHO line fitting the
profile with two Gaussians (see Table 1).
First contour and steps are, respectively, 5$\sigma$,
and 10$\sigma$, respectively.
The 1$\sigma$ value is: 7 mJy km s$^{-1}$ for 
SO$_2$(8$_{\rm 1,7}$--8$_{\rm 0,8}$ (Left panel),
14 mJy km s$^{-1}$ for SO$_2$(11$_{\rm 2,10}$--11$_{\rm 1,11}$),
SO$_2$(18$_{\rm 3,15}$--18$_{\rm 2,16}$), and
$^{34}$SO$_2$(12$_{\rm 0,12}$--11$_{\rm 1,11}$).
The filled ellipse shows the synthesised beam (HPBW): 1$\farcs$6 $\times$ 1$\farcs$1 (PA= 41$^\circ$) for the
SO$_2$(8$_{\rm 1,7}$--8$_{\rm 0,8}$) maps (Left panel), 
and 0$\farcs$65 $\times$ 0$\farcs$58 (PA= --46$^\circ$)
for the the other images. 
The black crosses indicate the positions of the VLA4A and VLA4B sources as imaged using the VLA array by \citet{Tobin2018}.
}
\end{figure*}

\begin{figure*}
\centerline{\includegraphics[angle=0,width=15cm]{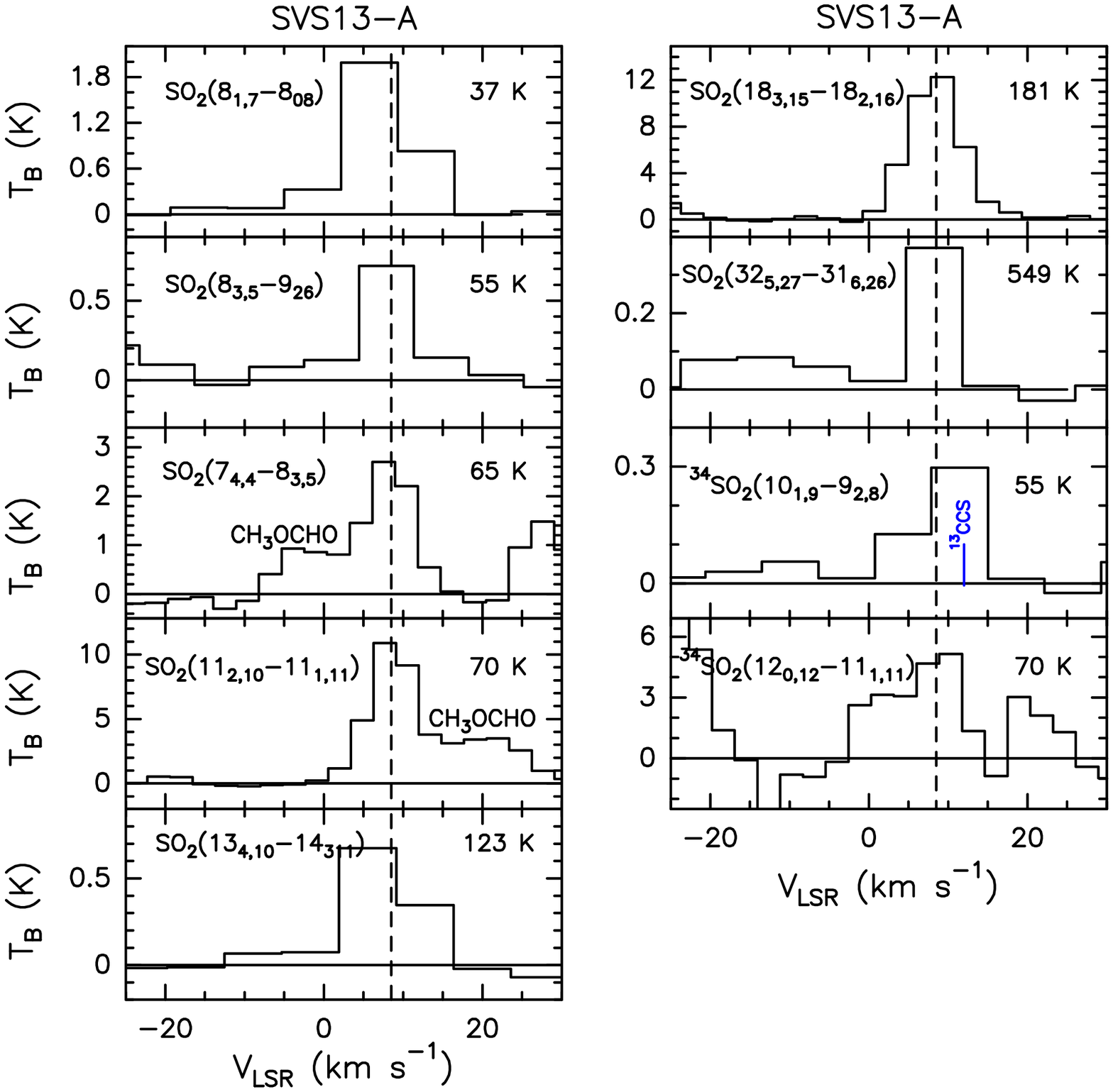}}
\caption{Observed SO$_2$, $^{34}$SO$_2$, and $^{33}$SO$_2$ spectra 
(in T$_{\rm B}$ scale, see Table 1) extracted at the emission peak: $\alpha_{2000}$ = 03$^h$29$^m$03$^s$.75,  $\delta_{2000}$ = +31$^\circ$16$'$03$''$.8. 
Transitions and corresponding upper level energies are reported.
The vertical dashed line stands for the ambient LSR velocity (+8.6 km s$^{-1}$, \citet{Chen2009}). The $^{34}$SO$_2$ line at $E_{\rm u}$ = 55 K
could be contaminated by $^{13}$CCS emission with $E_{\rm u}$ = 22 K
(see Table 1).}
\end{figure*}

\subsubsection{OCS}

Figure 8 shows the spatial distribution of the three detected OCS lines.
The $J$ = 7--6 and 8--7 transitions fall in the 3mm band: the emitting size 
is clearly unresolved with the spatial resolution of 1$\farcs$5.
Also the OCS(17--16) line, imaged with a synthesised beam 
of 0$\farcs$6, is spatially unresolved.
The $uv$ analysis leads to a peak coordinates of 
$\alpha_{2000}$ = 03$^h$29$^m$03$^s$.754,  
$\delta_{2000}$ = +31$^\circ$16$'$03$''$.778, with an error of 5 mas. 
This position is consistent with what found using
SO$_2$, lying between the VLA4A and VLA4B positions.
The OCS spectra are reported in Fig. 9: the peak velocity is 
in agreement with the 
SVS13-A systemic velocity of +8.6 km s$^{-1}$ \citep{Chen2009}. In addition
the OCS(17--16) profile, samples with a 2.8 km s$^{-1}$, shows a FWHM
of 7 km s$^{-1}$, consistent with those of the SO$_2$ lines observed at the same angular resolution.

\begin{figure}
\centerline{\includegraphics[angle=0,width=7cm]{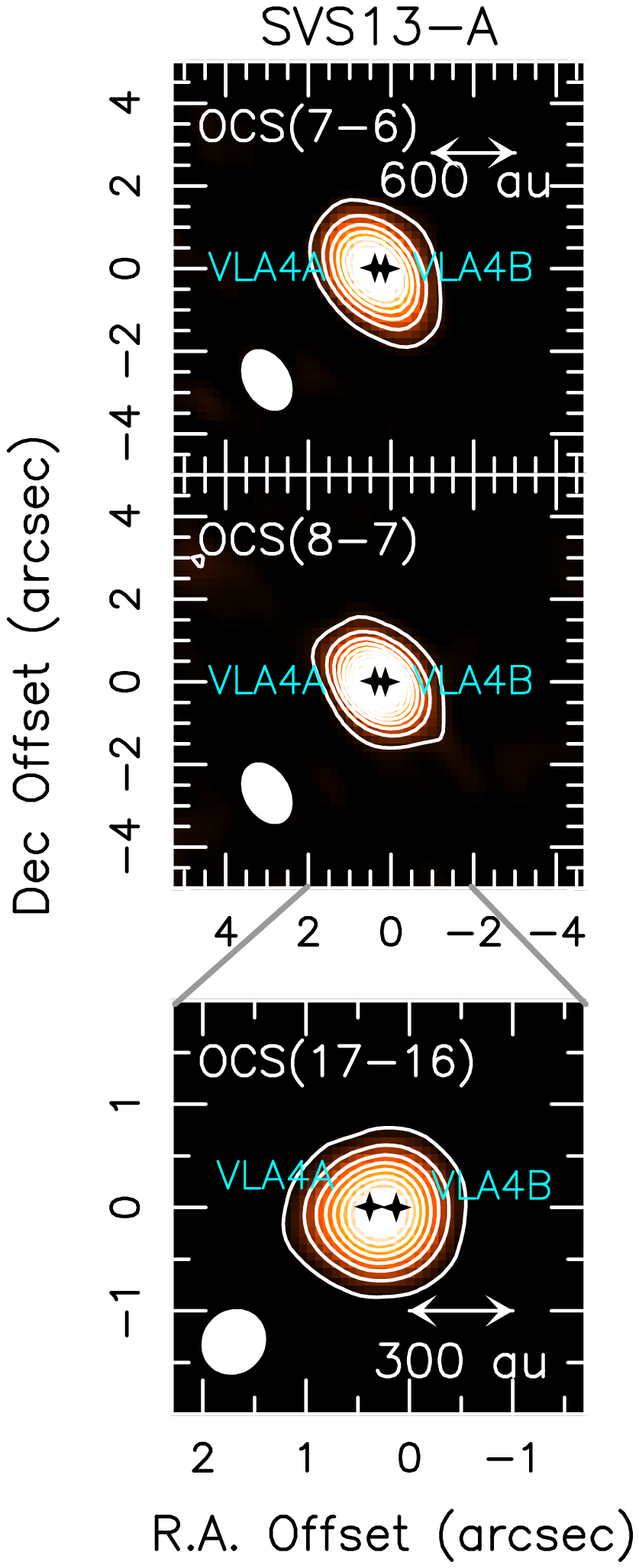}}
\caption{Spatial distribution of the
SVS13-A region obtained using the OCS(7--6) (Upper panel), OCS(8--7) (Middle),
and OCS(17--16) (Bottom) line emission (contours and colour). 
The whole velocity emitting region has been used (from 0 km s$^{-1}$
to +20 km s$^{-1}$, see Fig. 9).
Angular offset are with respect to the phase center 
(see Sect. 3).
First contour and steps are, respectively, 5$\sigma$,
and 10$\sigma$, respectively.
The 1$\sigma$ value is 5 mJy km s$^{-1}$ ($J$ = 7--6, and 8--7), and 
26 mJy km s$^{-1}$ ($J$ = 17--16). 
The filled ellipse shows the synthesised beam (HPBW): 1$\farcs$6 $\times$ 1$\farcs$1 (PA= 41$^\circ$) for
OCS(7--6) and OCS(8--7), and  
and 0$\farcs$65 $\times$ 0$\farcs$58 (PA= --46$^\circ$)
for OCS(17--16).
The black crosses indicate the positions of the VLA4A and VLA4B 
sources as imaged using the VLA array by \citet{Tobin2018}.}
\end{figure}

\begin{figure}
\centerline{\includegraphics[angle=0,width=8cm]{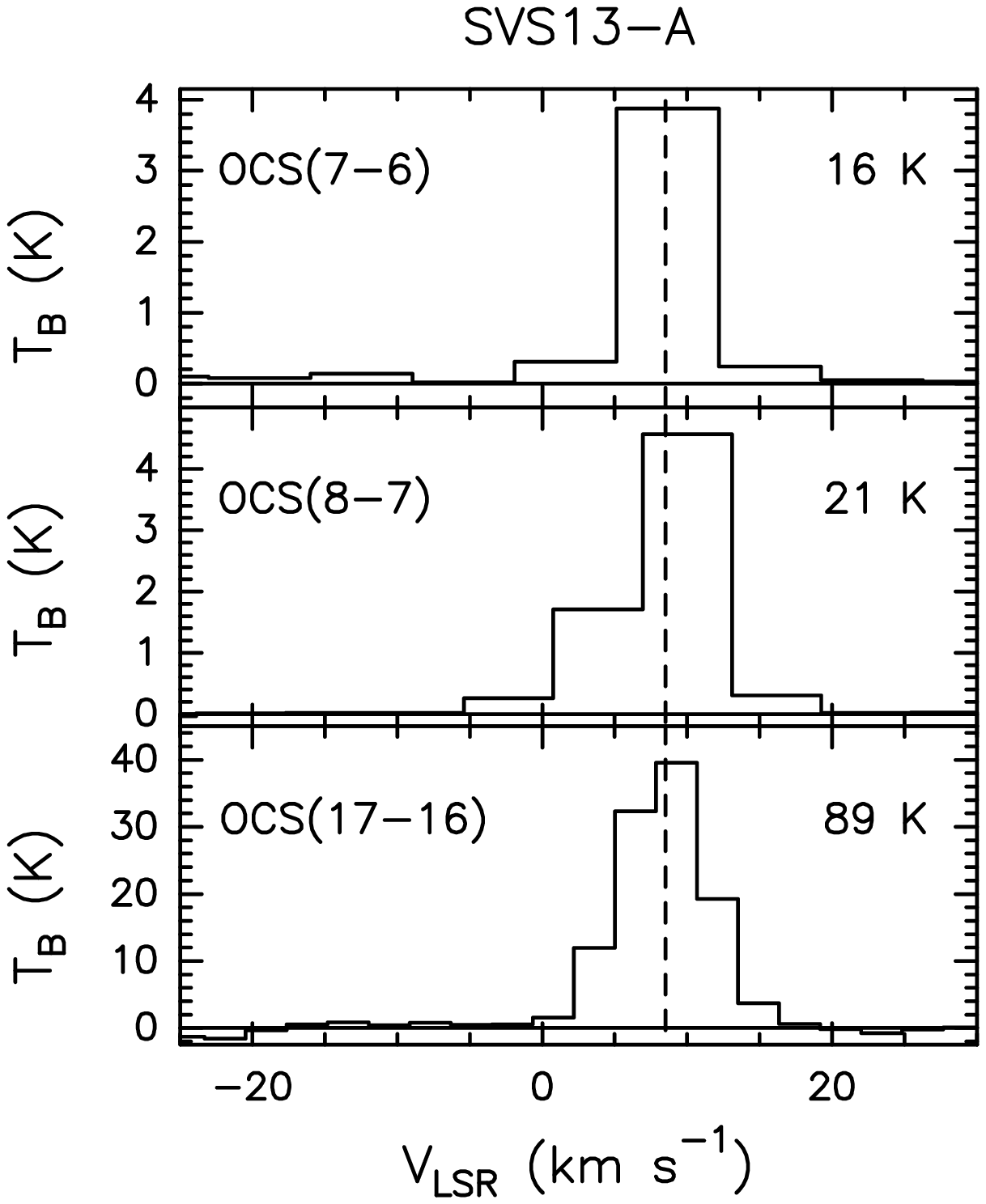}}
\caption{Observed OCS spectra 
(in T$_{\rm B}$ scale, see Table 1) extracted at the emission peak: $\alpha_{2000}$ = 03$^h$29$^m$03$^s$.75,  $\delta_{2000}$ = +31$^\circ$16$'$03$''$.8. 
Transitions and corresponding upper level energies are reported.
The vertical dashed line stands for the ambient LSR velocity (+8.6 km s$^{-1}$, \citet{Chen2009}).}
\end{figure}

\subsubsection{H$_2$CS}

As reported in Table 1, ten H$_2$CS and H$_2$CS$^{34}$S lines, 
with $E_{\rm u}$ in the 10--244 K range, 
have been revealed using both setups at 3mm and 1.4mm.
Two pairs of SO$_2$ lines are blended at the present spectral resolution.
Figure 10 shows examples of 
the spatial distributions of the H$_2$CS lines: 
as SO$_2$ and OCS, the emission is peaking towards SVS13-A,
being spatially unresolved at both 3mm and 1.4mm.

The peak coordinates,
according to the $uv$ fit, are 
$\alpha_{2000}$ = 03$^h$29$^m$03$^s$.752,  
$\delta_{2000}$ = +31$^\circ$16$'$03$''$.786, with an error of 8 mas,
in agreement with the SO$_2$ and OCS ones.
The profiles at the peak emission are shown in Fig. 11.
Interestingly, o-H$_2$C$^{34}$S(6$_{\rm 1,5}$--5$_{\rm 1,4}$)
has been observed with a spectral resolution of 0.9 km s$^{-1}$,
i.e. definitely better than that of the other lines (3--7 km s$^{-1}$).
For this line the profile is well sampled, peaking at the SVS13-A systemic
velocity (+8.6 km s$^{-1}$, \citealt{Chen2009}), and with
a FWHM of 2.7 km s$^{-1}$.

\begin{figure*}
\centerline{\includegraphics[angle=0,width=18cm]{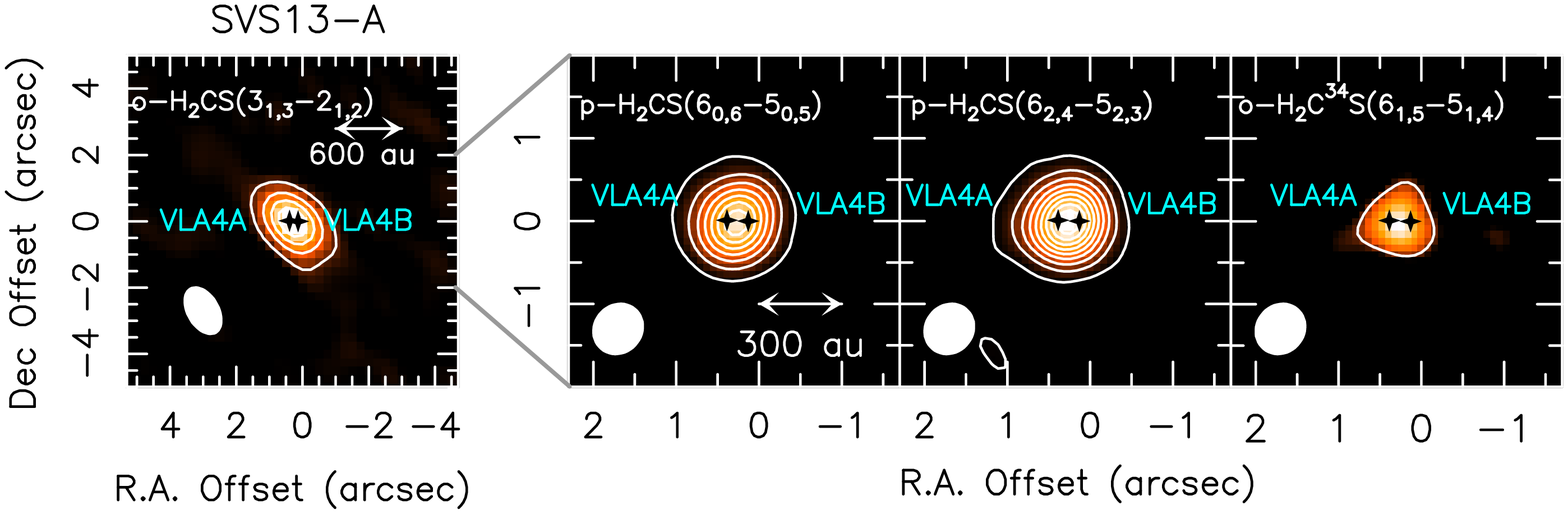}}
\caption{Spatial distribution (contours and colour) of the
SVS13-A region obtained using different H$_2$CS, and H$_2$C$^{34}$S line emission (see Table 1). 
The whole velocity emitting region has been used (from 0 km s$^{-1}$
to +20 km s$^{-1}$, see Fig. 11).
The p-H$_2$CS(6$_{\rm 2,4}$--5$_{\rm 2,3}$) emission has been deblended from the SO emission (see Table 1) applying two Gaussian profiles.
Angular offset are relative to the phase center
(see Sect. 2).
First contour and steps are, respectively, 5$\sigma$,
and 10$\sigma$, respectively.
The 1$\sigma$ value, from Left to Right, is 5, 13, 11, and 9 mJy km s$^{-1}$, respectively.
The filled ellipse shows the synthesised beam (HPBW): 1$\farcs$6 $\times$ 1$\farcs$1 (PA= 41$^\circ$) for
the o-H$_2$CS(3$_{\rm 1,3}$--2$_{\rm 1,2}$) 
line (Left panel), 
and 0$\farcs$65 $\times$ 0$\farcs$58 (PA= --46$^\circ$)
for the p-H$_2$CS(6$_{\rm 0,6}$--5$_{\rm 0,5}$),
p-H$_2$CS(6$_{\rm 2,4}$--5$_{\rm 2,3}$), and
o-H$_2$C$^{34}$S(6$_{\rm 1,5}$--5$_{\rm 1,4}$) lines (Middle and Right panels).
The black crosses indicate the positions of the VLA4A and VLA4B sources as imaged using the VLA array by \citet{Tobin2018}.}
\end{figure*}

\begin{figure*}
\centerline{\includegraphics[angle=0,width=15cm]{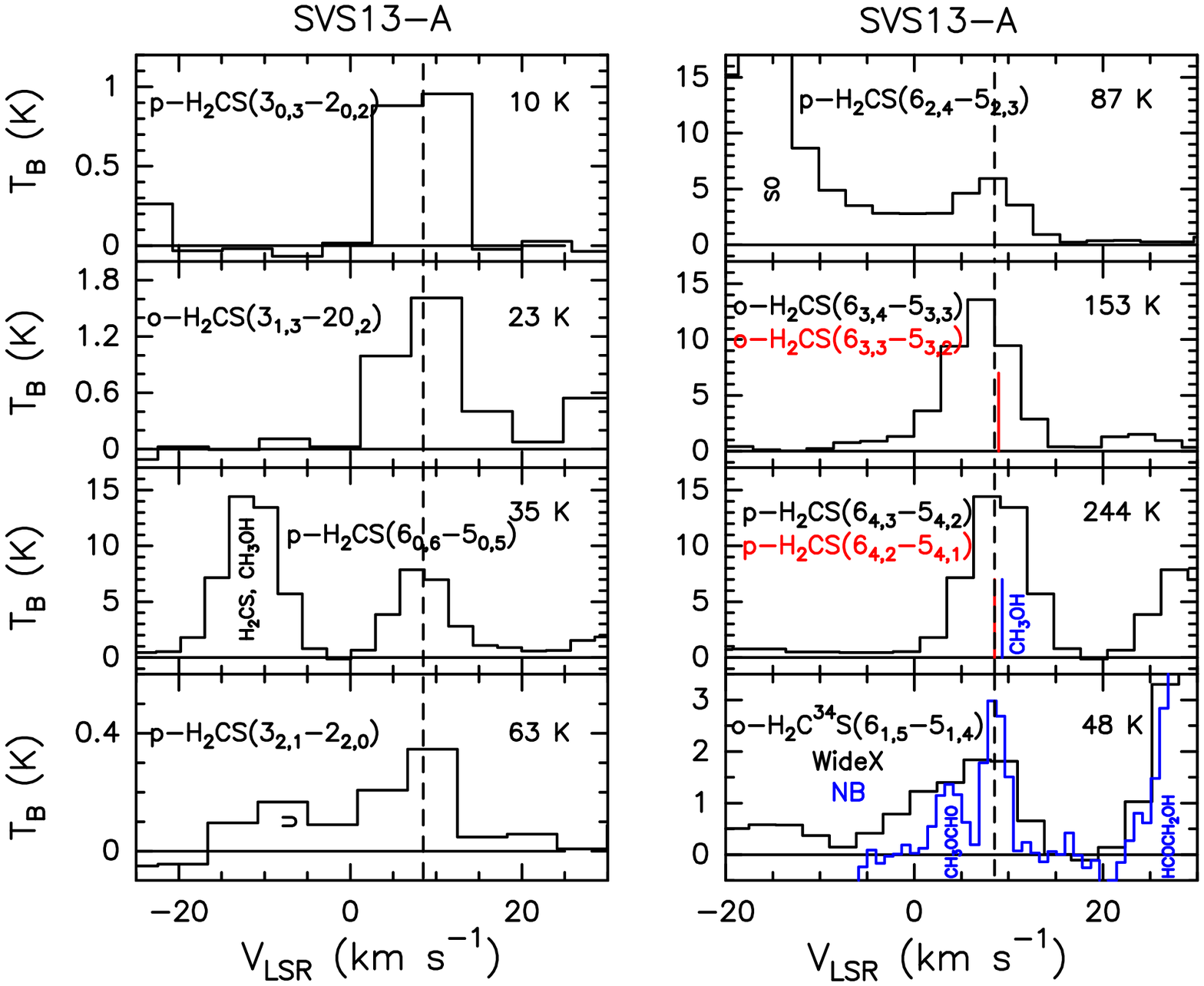}}
\caption{Observed H$_2$CS, and H$_2$C$^{34}$S spectra 
(in T$_{\rm B}$ scale, see Table 1) extracted at the emission peak: $\alpha_{2000}$ = 03$^h$29$^m$03$^s$.75,  $\delta_{2000}$ = +31$^\circ$16$'$03$''$.8. 
Transitions and corresponding upper level energies are reported.
The vertical dashed line stands for the ambient LSR velocity (+8.6 km s$^{-1}$, \citet{Chen2009}).
Red labels and red vertical segments denote two H$_2$CS profiles
with the same $E_{\rm u}$ value and blended at the
present spectral resolution (Table 1). The black vertical 
labels indicate that the line is reported in another panel
of the present figure. The H$_2$CS doublets at $E_{\rm u}$ = 244 K
could be contaminated by CH$_3$OH emission with $E_{\rm u}$ = 317 K
(see Table 1). The H$_2$C$^{34}$S(6$_{1,5}$--5$_{1,4}$) profile has
been observed using both low- (black histogram) and high-spectral (blue) backends (see Sect. 2: labelled Widex and NB, respectively).}
\end{figure*}

\subsubsection{NS}

The NS(9/2--7/2) $\Omega$=1/2 line, with $E_{\rm u}$ = 27 K, has been detected
towards SVS13-A. The profile consists of three hyperfine components,
blended with the 2.8 km s$^{-1}$ spectral resolution (see Fig. 5). 
The frequency transitions is 207.4 GHz and the corresponding spatial 
distribution is shown in Fig. 12. The emitting size is unresolved, being less than the 0$\farcs$6 synthesised beam. The emission peaks between
the positions of the VLA4A and VLA4B, namely at
$\alpha_{2000}$ = 03$^h$29$^m$03$^s$.750,  
$\delta_{2000}$ = +31$^\circ$16$'$03$''$.808, with an error of 14 mas.
In summary, NS, OCS, SO$_2$, H$_2$CS, once observed with the present spatial resolution ($\sim$ 100 au) are tracing the same spatially unresolved region,
being consistent with what expected from a chemical enriched region around the protostar(s), i.e. the hot corino(s). 
These findings will be discussed in Sect. 6 in light of the physical parameters 
derived in Sect. 5.

\begin{figure}
\centerline{\includegraphics[angle=0,width=6cm]{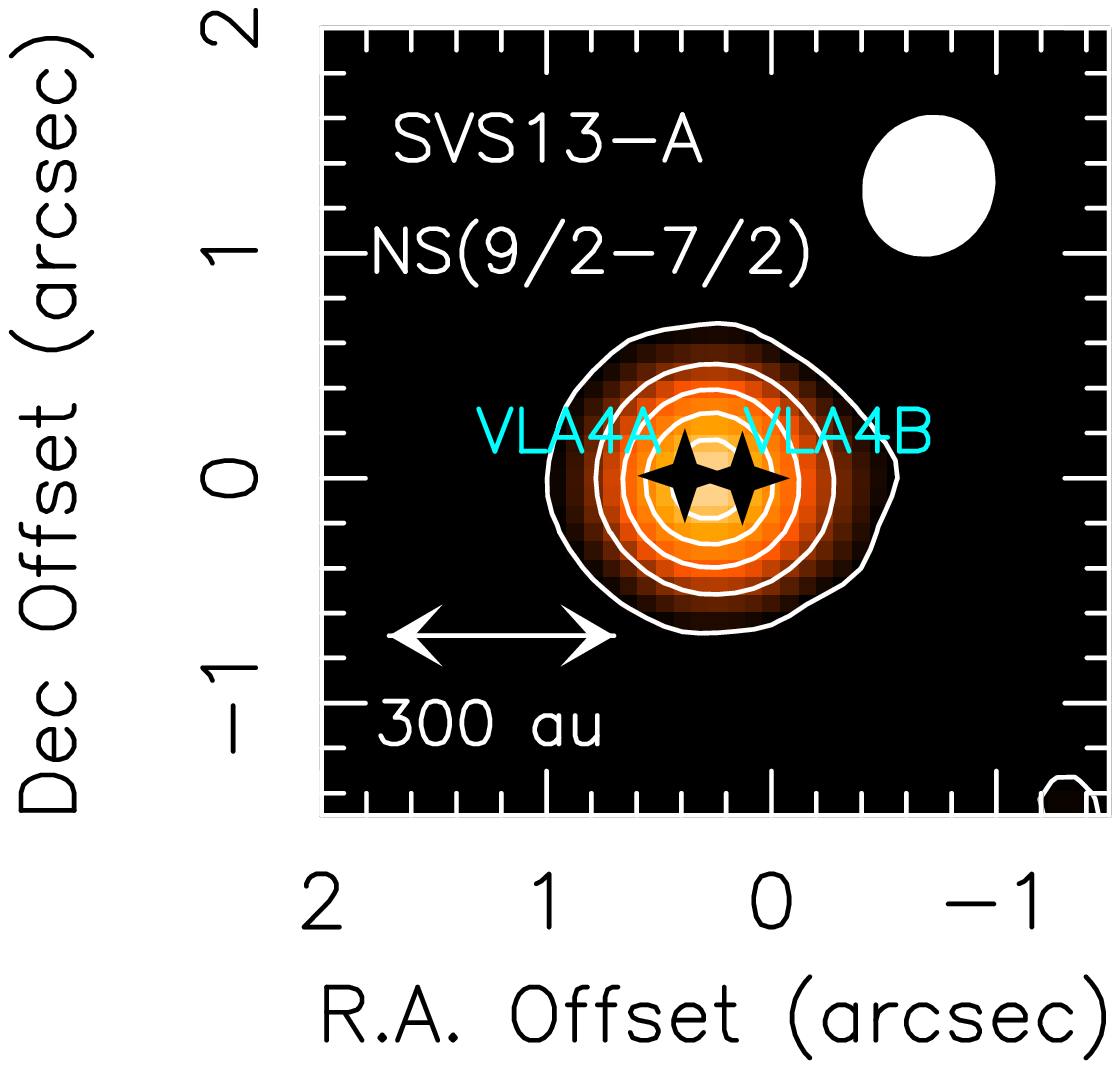}}
\caption{Spatial distribution (contours and colour) of the
SVS13-A region obtained using the NS(9/2--7/2) $\Omega$=1/2 emission. 
The whole velocity emitting region has been used (from 0 km s$^{-1}$
to +15 km s$^{-1}$. The line profile consists of three hyperfine components, blended
at the present spectral resolution (see Fig. 5 and Table 1).
Angular offset are relative to the phase center
(see Sect. 3).
First contour and steps are, respectively, 5$\sigma$
(50 mJy km s$^{-1}$),
and 10$\sigma$, respectively.
The filled ellipse shows the synthesised beam (HPBW): 0$\farcs$65 $\times$ 0$\farcs$58 (PA= --46$^\circ$).
The black crosses indicate the positions of the VLA4A and VLA4B 
sources as imaged using the VLA array by \citet{Tobin2018}.}
\end{figure}

\begin{table*}
\caption{List of transitions of S-bearing species detected towards SVS13-A. The Gaussian fit parameters (in $T_{\rm MB}$ scale) refer to the spectra extracted at the emission peak: $\alpha_{2000}$ = 03$^h$29$^m$03$^s$.75,  $\delta_{2000}$ = +31$^\circ$16$'$03$''$.8.}
\begin{tabular}{lccccccccc}
 \hline
\multicolumn{1}{c}{Transition} &
\multicolumn{1}{c}{$\nu$$^{\rm a}$} &
\multicolumn{1}{c}{$E_{\rm up}$$^a$} &
\multicolumn{1}{c}{$S\mu^2$$^a$} &
\multicolumn{1}{c}{$d$v} &
\multicolumn{1}{c}{rms} &
\multicolumn{1}{c}{$T_{\rm peak}$} &
\multicolumn{1}{c}{$V_{\rm peak}$} &
\multicolumn{1}{c}{FWHM} &
\multicolumn{1}{c}{$I_{\rm int}$} \\
\multicolumn{1}{c}{ } &
\multicolumn{1}{c}{(MHz)} &
\multicolumn{1}{c}{(K)} &
\multicolumn{1}{c}{(D$^2$)} & 
\multicolumn{1}{c}{(km s$^{-1}$)} &
\multicolumn{1}{c}{(mK)} &
\multicolumn{1}{c}{(mK)} &
\multicolumn{1}{c}{(km s$^{-1}$)} &
\multicolumn{1}{c}{(km s$^{-1}$)} &
\multicolumn{1}{c}{(K km s$^{-1}$)} \\ 
\hline
\multicolumn{10}{c}{SO}\\
2$_{\rm 2}$--1$_{\rm 1}$ & 86093.95 & 19  & 3.5  & 7.0 & 22 & 2765(98) &  +8.9(0.2) & 9.7(0.5) & 27.69(0.23) \\
2$_{\rm 3}$--1$_{\rm 2}$ & 99299.87 & 9 & 6.9 & 6.0 & 42 & 4191(42) & +8.7(0.2) &  9.4(0.7) & 44.30(1.99) \\
4$_{\rm 5}$--4$_{\rm 4}$ & 100029.64 & 39 & 0.8 & 6.0 & 23 & 1209(43) &  +8.8(0.2) & 7.5(0.5) & 9.55(0.27) \\
4$_{\rm 5}$--3$_{\rm 4}$ & 206176.01 & 39 & 8.9 & 2.8 & 920 & 43142(777) & +8.3(0.2) & 9.2(0.2) & 422.00(7.38) \\

\hline
\multicolumn{10}{c}{$^{34}$SO}\\
2$_{\rm 2}$--1$_{\rm 1}$ & 84410.69 & 19 & 3.5 & 7.1 & 54 & 296(12) & +9.2(4.2)  & 7.2(5.1) & 2.29(0.68) \\
2$_{\rm 3}$--1$_{\rm 2}$ & 97715.32 & 9 & 6.9 & 6.1 & 120 & 436(4) & +8.2(1.1) & 9.7(7.0) & 4.48(1.20) \\

\hline
\multicolumn{10}{c}{SO$_2$}\\
13$_{\rm 4,10}$--14$_{\rm 3,11}$ & 82951.94  & 123 & 5.1 & 7.1 & 57 & 751(23) & +7.6(1.7) & 9.9(1.5) & 7.96(0.68) \\
8$_{\rm 1,7}$--8$_{\rm 0,8}$ & 83688.09   & 37 & 17.0 &  7.2 & 72 & 2071(24) & +7.0(0.3) & 10.3(0.5) & 22.65(0.94) \\
32$_{\rm 5,27}$--31$_{\rm 6,26}$ & 84320.88  & 549 & 13.5 & 7.1 & 40 & 364(30) & +8.0(2.3) & 7.1(6.1) & 2.76(0.65) \\
8$_{\rm 3,5}$--9$_{\rm 2,6}$ & 86639.09 & 55  & 3.0 & 6.9 & 53 & 719(22) & +8.1(1.3) & 8.9(1.4) & 6.82(0.73) \\
18$_{\rm 3,15}$--18$_{\rm 2,16}$ & 204246.76  &  181 & 34.5 & 2.9 & 780 & 12609(160) & +8.2(0.2) & 7.6(0.5) & 102.64(5.5) \\
7$_{\rm 4,4}$--8$_{\rm 3,5}$ & 204384.30  & 65 & 1.7 & 2.9 &  540 &  2757(210) & +8.0(0.7) & 7.9(1.6) & 23.09(4.12) \\
11$_{\rm 2,10}$--11$_{\rm 1,11}$ & 205300.57  & 70 & 12.1 & 2.9 & 197 & 11235(230) & +8.7(0.8) & 7.3(1.9) & 87.28(19.24) \\

\hline
\multicolumn{10}{c}{$^{34}$SO$_2$}\\
10$_{\rm 1,9}$--9$_{\rm 2,8}$$^b$ & 82124.35  & 55 & 6.5 & 7.1 & 160 & 6998(55) & +9.4(1.2) & 8.3(3.2) & 63.29(3.03) \\
12$_{\rm 0,12}$--11$_{\rm 1,11}$ & 204136.23  & 70 & 22.7 & 2.9 & 1400 & 5982(780) & +9.1(0.7) & 5.4(2.0)  & 34.56 (9.4) \\


\hline
\multicolumn{10}{c}{CS}\\
2--1 & 97980.95 & 7 & 7.6 & 6.1 & 38 & 5196(12) & +8.6(0.2) & 7.0(0.3) & 38.72(0.62) \\

\hline
\multicolumn{10}{c}{C$^{34}$S}\\
2--1 & 96412.95 & 6 & 7.6 & 6.2 & 45 & 2088(27) & +7.8(1.4) & 6.4(4.8) & 14.26(4.38) \\

\hline
\multicolumn{10}{c}{C$^{33}$S}\\
2--1 $F$=5/2--3/2$^c$ & 97171.84 & 7 & 12.2 & 6.2 & 29 & 388(32) & +8.2(0.5) & 8.6(1.2) & 3.56(0.33) \\

\hline
\multicolumn{10}{c}{OCS}\\
7--6 & 85139.10 & 16 & 3.5 & 7.0 & 36 & 3787(50) & +8.5(0.4) & 7.1(0.2) & 28.39(0.34) \\
8--7 & 97301.21 & 21 & 4.1 & 6.1 & 29 & 5031(86) & +8.6(0.2) & 7.6(0.2) & 
40.73(1.18) \\
17--16 & 206745.16 & 89 & 8.7 & 2.8 & 471 & 40764(140) & +8.4(0.3)  & 7.1(0.7)  & 308.66(4.89) \\


\hline
\multicolumn{10}{c}{H$_2$CS}\\
o-3$_{\rm 1,3}$--2$_{\rm 1,2}$ & 101477.81 & 23 & 21.8 & 5.9 & 20 & 1576(12) & +8.6(1.4)  & 10.0(3.3) & 17.12(4.50) \\
p-3$_{\rm 0,3}$--2$_{\rm 0,2}$ & 103040.45 & 10 & 8.2 & 5.8 & 33 & 1603(93) & +8.5(1.3)  & 6.2(3.3) & 10.49(2.26) \\
p-3$_{\rm 2,1}$--2$_{\rm 2,0}$ & 103051.87 & 63 & 4.5 & 5.8 & 46 & 318(18) & +7.8(4.7)  & 9.9(4.5) & 3.79(1.15) \\
p--6$_{\rm 0,6}$--5$_{\rm 0,5}$ & 205987.86 & 35 & 16.3 & 2.8 & 159 & 7702(79) & +8.2(0.2) & 7.0(0.2) & 57.35(0.56) \\
p-6$_{\rm 4,3}$--5$_{\rm 4,2}$$^d$ & 206001.88$^d$ & 244 & 9.0 & & & & \\ 
 & & & & \Big\}  2.8 & 115 & 15074(260) & +8.8(0.2) & 7.3(0.2) & 116.65(1.84) \\
p-6$_{\rm 4,2}$--5$_{\rm 4,1}$$^d$ & 206001.88$^d$ & 244 & 9.0 & &  &  & &  & \\
o-6$_{\rm 3,4}$--5$_{\rm 3,3}$$^d$ & 206051.94$^d$ & 153 & 36.6  & &  &  &  & & \\
 &  &  &   & \Big\}  2.8 & 161 & 13512(120) & +7.1(0.8) & 7.6(2.1) & 108.84(24.75) \\ 
o-6$_{\rm 3,3}$--5$_{\rm 3,2}$$^d$ & 206052.24$^d$ & 153 & 36.6 & &  &  & &  & \\
p-6$_{\rm 2,4}$--5$_{\rm 2,3}$$^e$ & 206158.60$^e$ & 87 & 14.5 & 2.8 & 160  & 6004(120) & +7.9(0.2) & 7.7(0.5) & 48.87(0.49) \\

\hline
\multicolumn{10}{c}{H$_2$C$^{34}$S}\\
o-6$_{\rm 1,5}$--5$_{\rm 1,4}$ & 205583.13 & 48 & 47.5 & 0.9 & 560 & 3113(190) & +8.6(0.2) & 2.7(0.5) & 8.91(1.36) \\

\hline
\multicolumn{10}{c}{NS}\\
9/2--7/2 $\Omega$=1/2$^f$ l=e & 207436.05 & 27 & 17.4 & 2.8 & 1100 & 3968(840) & +7.9(0.8) & 5.7(1.7) & 24.22(6.41) \\

\hline
\multicolumn{10}{c}{NS$^+$}\\
2--1 & 100198.55 & 7 & 8.7 & 6.0 & 28 & -- & -- & -- & $\leq$ 1$^g$ \\

\hline
\vspace{2mm}
\end{tabular}

$^a$ Spectral parameters from the Cologne Database for Molecular Spectroscopy \citep{Muller2001,Muller2005} for all the species, with the exception of
those of the H$_2$CS isotoplogues, derived from the Jet Propulsion Laboratory 
\citep[JPL][]{Pickett1998}, and those of NS$^+$, from \citet{Cernicharo2018}.
$^b$ Possible contamination with $^{13}$CCS emission at 82123.376 MHz ($E_{\rm u}$ = 22 K, $S\mu^2$ = 0.7 D$^2$).
$^c$ The C$^{33}$S(2--1) line consists of 6 hyperfine
components \citep{Bogay1981,Lovas2004,Muller2005} in a 9 MHz frequency interval. The line with the highest $S\mu^2$
is reported (see Fig. 5).
$^d$ Lines blended at the present spectral resolution (2 MHz). Possible contamination with CH$_3$OH emission at 206001.30 MHz ($E_{\rm u}$ = 317 K, $S\mu^2$ = 10.2 D$^2$).
$^e$ Contaminated by SO(4$_{\rm 5}$--3$_{\rm 4}$) high-velocity emission.
$^f$ The 9/2--7/2 $\Omega$=1/2 line consists of 3 hyperfine
components \citep{Lee1995-NS} in a 0.6 MHz frequency interval. The line with the highest $S\mu^2$ $F$=11/2--9/2
is reported (see Fig. 5).
$^g$ 3$\sigma$ upper limit.
\label{table:lines}
\end{table*}

\section{Physical parameters}\label{sec-phys-parameters}

We analysed the SO, SO$_2$ and H$_2$CS (and their isotopologues) spectra extracted at the emission peak 
($\alpha_{2000}$ = 03$^h$29$^m$03$^s$.75,  $\delta_{2000}$ = +31$^\circ$16$'$03$''$.8) via the non-LTE (Local Thermodynamic Equilibrium) Large Velocity Gradient (LVG) approach using the code \texttt{grelvg} described in \citet{Ceccarelli2003}.  
We assumed a Boltzmann distribution for the H$_2$ ortho-to-para
ratio. For SO, we used the collisional coefficients with p-H$_{2}$ computed by \citet{Lique2007} and retrieved from the BASECOL database \citep{Dubernet2013}.
For SO$_2$, we used the collisional coefficients with ortho- and 
para-H$_{2}$ computed by \cite{balanca2016} and retrieved from the LAMDA database \citep{Schoier2005}. 
For H$_2$CS, we used the H$_2$CO-H$_{2}$ collisional coefficients with ortho- and para-H$_{2}$ derived by \citet{Wiesenfeld2013}, scaled them for the mass ratio, and provided by the LAMDA database \citep{Schoier2005}. 

We assumed a semi-infinite slab geometry to compute the line escape probability \citep{Scoville1974} and assumed a line width equal to 8 km s$^{-1}$.
We used the $^{32}$S/$^{34}$S equal to 22 \citep{WilsonRood}, and an H$_2$CS ortho-to-para ratio equal to 3.
Finally, the errors on the observed line intensities have been obtained adding the spectral r.m.s. to the uncertainties due to calibration (see Sect. 3).
We ran large grids of models varying the kinetic temperature ($T_{\rm kin}$) from 20 to 500 K, the volume density ($n_{\rm H_2}$) from 10$^{4}$ cm$^{-3}$ to 10$^{9}$ cm$^{-3}$, the emitting sizes from 0$\farcs$1 to 30$\arcsec$ and the
SO, SO$_2$ and H$_2$CS column densities from 10$^{15}$ cm$^{-2}$ 
to 10$^{21}$ cm$^{-2}$. 
The best fit for each species is obtained by simultaneously minimizing the difference between the predicted and observed integrated intensities of both used isotopologues (e.g. SO and $^{34}$SO; SO$_2$ and $^{34}$SO$_2$; H$_2$CS and H$_2$C$^{34}$S).


The results obtained for SO show a best solution, characterised by $\chi^{2}$ $\simeq$ 2, with a total column density 
N$_{\rm SO}$ of 10$^{17}$ cm$^{-2}$, a
kinetic temperature $T_{\rm kin}$ equal to 280 K,
and the volume density $n_{\rm H_2}$ being 10$^{8}$ cm$^{-3}$.
The size is 0$\farcs$3 (90 au). 
The line opacities lies between 0.1 and 1.4.
If we consider an uncertainty of 1 $\sigma$, which correspond to 
a probability of 30\% to exceeding $\chi^{2}$, we have
$T_{\rm kin}$ $\geq$ 150 K, $n_{\rm H_2}$ $\geq$ 6 $\times$ 10$^{6}$ cm$^{-3}$,
N$_{\rm SO}$ = 0.2--3 $\times$ 10$^{17}$ cm$^{-2}$, and sizes in the 0$\farcs$2--0$\farcs$5 range. 

The analysis of SO$_2$ leads to a best solution 
($\chi^{2}$ $\simeq$ 3) with 
the N$_{\rm SO_2}$ = 10$^{18}$ cm$^{-2}$,
$T_{\rm kin}$ = 210 K, and  10$^{8}$ cm$^{-3}$.
The size is 0$\farcs$2 ($\sim$ 60 au).
The line emission is moderately thick, being the
opacity between 0.30 and 4.5.
Once considered the uncertainty of 1 $\sigma$, we obtain
$T_{\rm kin}$ $\simeq$ 100--300 K, and $n_{\rm H_2}$ $\geq$ 5 $\times$ 10$^{6}$ cm$^{-3}$, N$_{\rm SO_2}$ = 0.3--3 $\times$ 10$^{18}$ cm$^{-2}$, and the size between 0$\farcs$1 and 0$\farcs$3. 

The best fit obtained for H$_2$CS, identified by
$\chi^{2}$ $\simeq$ 0.1, indicates the total
column density equal to 2 $\times$ 10$^{15}$ cm$^{-2}$, 
a size of 0$\farcs$4 (120 au), the kinetic
temperature $T_{\rm kin}$ = 100 K, and the
volume density $n_{\rm H_2}$ = $2 \times$ 10$^{5}$ cm$^{-3}$.
The line opacities are all in the 0.04--0.4 range but that
of the 6$_{0,6}$--5$_{0,5}$ para line which is 0.86.
Taking into account 1 $\sigma$, the following constraints are
derived: $T_{\rm kin}$ $\geq$ 50 K, $n_{\rm H_2}$ $\geq$ 10$^{5}$ cm$^{-3}$,
N$_{\rm H_2CS}$ = 0.7--2 $\times$ 10$^{15}$ cm$^{-2}$, and sizes in the 0$\farcs$2--0$\farcs$8 range. 

Figure 13 shows the comparison between observations and 
best-fit line predictions for SO, SO$_2$, and H$_2$CS,
while Table 2 summarises the excitation ranges found for the three molecules. As a matter of fact, all the SO, SO$_2$ and H$_2$CS analysis are consistent with the occurrence of hot corino emission. This will allow us to compare the present results with what previously
found using iCOMs (see Sect. 6). 

\begin{table*}
 \caption{1$\sigma$ Confidence Level (range) from the Non-LTE LVG Analysis of the SO, SO$_2$, and H$_2$CS lines towards SVS13-A as imaged with NOEMA. The  OCS, CS, and NS column densities, derived assuming LTE conditions, are also reported. The last row with H$_2$S, is based on IRAM 30-m data at 1.4mm. The last column is for the abundance
 with respect to H$_2$.}
    \begin{tabular}{c|ccccccc}
         \hline
         Species & N$_{\rm tot}$ & n$_{\rm H_2}$ & T$_{\rm kin}$ & size & X$_{\rm H_2}$$^d$ \\
          & (cm$^{-2}$) & (cm$^{-3}$) & (K) & (arcsec) &  \\
          \hline
          SO & 
                     0.2-3 $\times$ 10$^{17}$ & $\geq$ 6 $\times$ 10$^{6}$ & $\geq$ 150 K & 0$\farcs$2--0$\farcs$5 &  7 $\times$ 10$^{-9}$ -- 1 $\times$ 10$^{-7}$ \\
          SO$_2$ & 0.3-3 $\times$ 10$^{18}$ & $\geq$ 5 $\times$ 10$^{6}$ & 100--300 &   0$\farcs$1--0$\farcs$3 & 1--10 $\times$ 10$^{-7}$ \\
          H$_2$CS  & 0.7-2 $\times$ 10$^{15}$ & $\geq$ 10$^{5}$  & $\geq$ 50 K & 0$\farcs$2--0$\farcs$8 & 2--7 $\times$ 10$^{-10}$ \\
          \hline
 & N$_{\rm tot}$ & n$_{\rm H_2}$ & T$_{\rm rot}$ & size & X$_{\rm H_2}$$^d$ \\
           & (cm$^{-2}$) & (cm$^{-3}$) & (K) & (arcsec)  &   \\
          \hline
         OCS$^a$ & 1-2 $\times$ 10$^{15}$ & -- & 70--170 &   0$\farcs$3$^a$ & 3--7 $\times$ 10$^{-10}$ \\
         CS$^b$ & 0.8-17 $\times$ 10$^{18}$ & -- & 37.5$^c$--300 &   0$\farcs$3$^b$ & 0.3--6 $\times$ 10$^{-6}$ \\
         NS$^b$ & 1-6 $\times$ 10$^{15}$ & -- & 37.5$^c$--300 &  0$\farcs$3$^b$ & 0.2--3 $\times$ 10$^{-9}$ \\
         NS$^+$$^b$ & $\leq$ 8 $\times$ 10$^{14}$ & -- & 37.5$^c$--300 &  0$\farcs$3$^b$ & $\leq$ 3 $\times$ 10$^{-10}$ \\
         H$_2$S$^{b,e}$ & 0.5--1 $\times$ 10$^{18}$ & -- & 37.5$^c$--300 & 
         0$\farcs$3$^b$ & 2--4 $\times$ 10$^{-7}$ \\
         \hline
    \end{tabular}
    
    $^a$ Estimate derived using the RD approach, and assuming a source size of
    0$\farcs$3 following the SO, SO$_2$, and H$_2$CS LVG analysis. 
    $^b$ Estimates derived using one line, and assuming: (i) a source size of
    0$\farcs$3, and a temperature in the 37.5--300 K range (from
    LVG analysis of SO, SO$_2$, and H$_2$CS). For CS, we took into account 
    the line opacity of 14 derived from the comparison with the emission of 
    C$^{34}$S and C$^{33}$S.
    $^c$ For T$_{\rm rot}$ we conservatively used a lower limit of 37.5 K, according to the partition function tabulated in the Jet Propulsion Laboratory \citep[JPL,][]{Pickett1998} database. 
    $^d$ We assume N$_{\rm H_2}$= 3 $\times$ 10$^{24}$ cm$^{-2}$ \citep{Chen2009}.
    $^e$ Estimated based on the 2$_{2,0}$--2$_{1,1}$ line observed
    with a HPBW = 11$\arcsec$ using the IRAM 30-m antenna (see Sect. 6.4).
    \label{tab:LVG_results}
\end{table*}

\begin{figure}
\centerline{\includegraphics[angle=0,width=10cm]{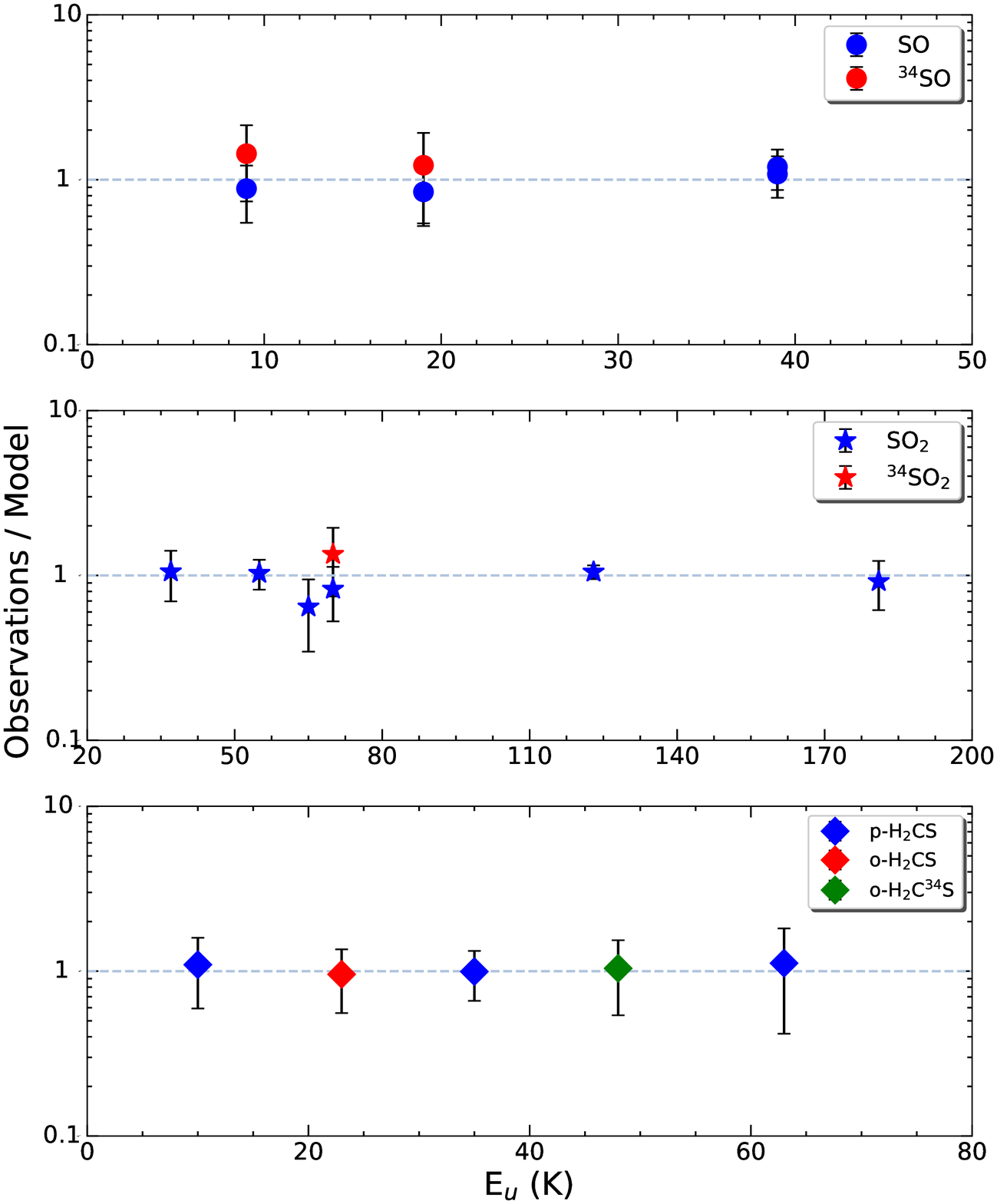}}
\caption{Ratio of the observed and best-fit theoretical line integrated intensities of
SO (Upper panel), SO$_2$ (Middle), and H$_2$CS (Bottom) as a function of the line upper-level energy (see the text and Table 2). 
Different symbols indicate different isotopologues as well as ortho and para species.}
\end{figure}

For OCS, having detected three lines, two of them at very similar upper level excitation (16 K and 21 K), we did not apply the LVG analysis, adopting instead 
the rotational diagram (RD) approach,
where LTE population and optically thin lines are assumed.
Under these assumptions, for a given molecule, the relative population
distribution of all the energy levels, is described by a Boltzmann
temperature, that is the rotational temperature $T_{\rm rot}$.
Given the OCS emission is peaking towards SVS13-A and spatially unresolved
(Fig. 8), we corrected the line intensities assuming a size of 0$\farcs$3, i.e.
the average size obtained from the LVG results for SO, SO$_2$, and H$_2$CS.
The RD analysis which provides a column density of
1--2 $\times 10^{15}$ cm$^{-2}$, and a rotational temperature 
T$_{\rm rot}$ = 120$\pm$50 K (see Fig. 14). 

Finally, for CS and NS, having detected only one transition, in light of the 
previous LVG results we adopted again a source size of 0$\farcs$3 and conservatively 
assumed the overall temperature range (37.5--300 K, see Table 2).
We consequently derived: 0.8--17 $\times 10^{18}$ cm$^{-2}$ for CS, and
1--6 $\times 10^{15}$ cm$^{-2}$ for NS.

\begin{figure}
\centerline{\includegraphics[angle=0,width=9.5cm]{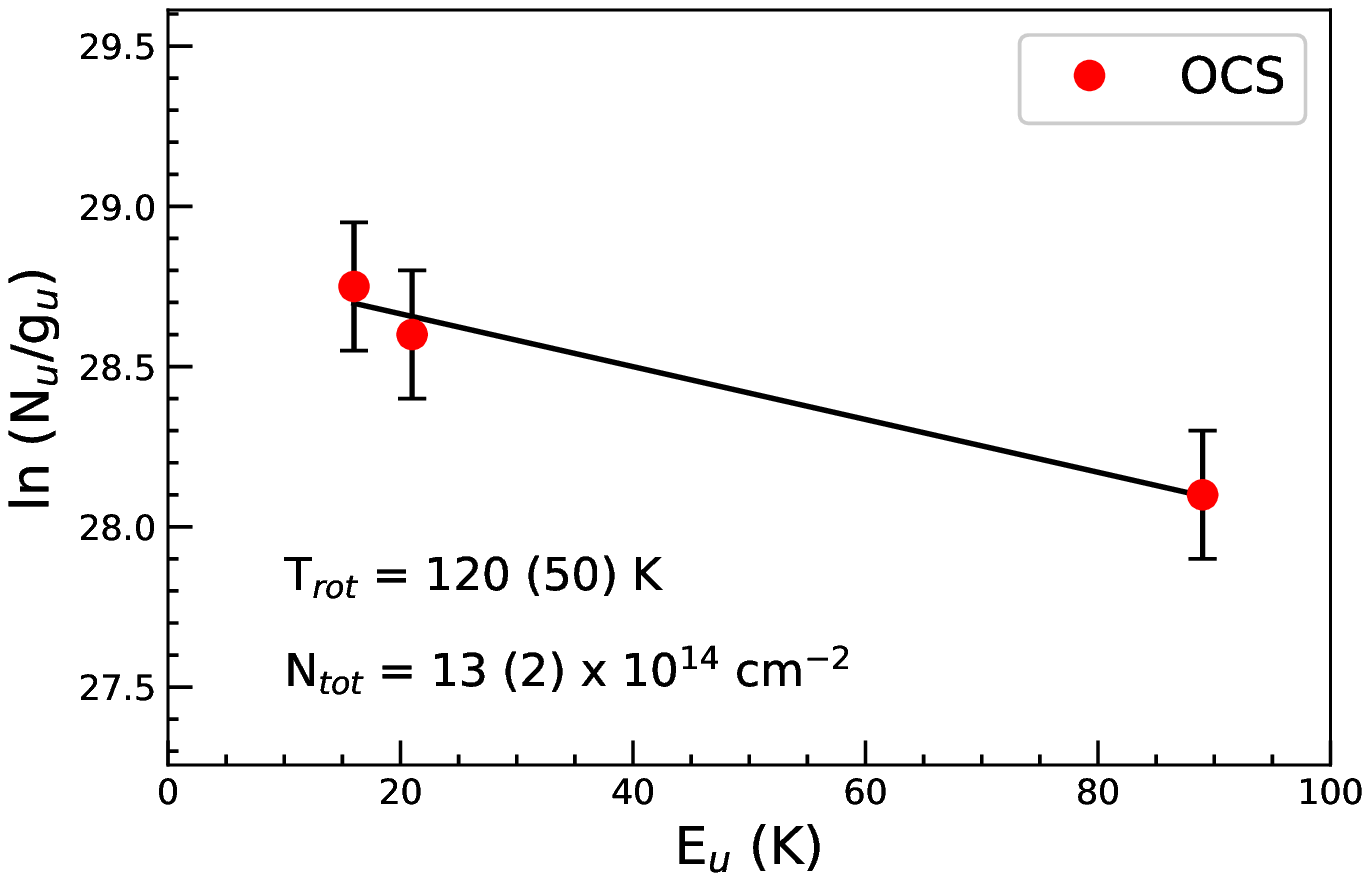}} 
\caption{Rotational diagrams for OCS derived using the emission lines
  observed towards SVS13-A (see Table 1 and Fig. 9).  The parameters
  N$_{u}$, g$_{u}$, and $E_{\rm up}$ are, respectively, the column
  density, the degeneracy and the energy (with respect to the ground
  state of each symmetry) of the upper level. A source size of
  0$\farcs$3 has been assumed, following the SO, SO$_2$, and H$_2$CS
  LVG analysis. The derived values of
  the rotational temperature are reported in the panels.}
\label{RD-OCS}
\end{figure}

\section{Discussion}\label{sec:discussion}

\subsection{The chemical census of the SVS13-A hot corino}

The SVS13-A system has been recently subject of IRAM 30-m 
\citep[ASAI:][]{Lefloch2018} and IRAM PdBI 
\citep[CALYPSO:][]{Maury2019} observations aimed to obtain its chemical census. More specifically, \citet{Bianchi2017a,Bianchi2019a} 
reported the analysis of a large sample of iCOMs using the ASAI single dish unbiased spectral survey: the large number of lines allowed the authors to: (i) analyse using the LVG approach the methanol
isotopologues, and (ii) consequently derive the column
densities of more complex organic species
(CH$_3$CHO, H$_2$CCO, HCOOCH$_3$, CH$_3$OCH$_3$, CH$_3$CH$_2$OH, NH$_2$CHO).
On the other hand, \citet{Desimone2017} and \citet{Belloche2020}
used the PdBI images (synthesised beams between 0$\farcs$5 and
1$\farcs$7) of iCOMs emitting in selected spectral
windows to derive the column densities in LTE conditions
of, again, a large number of iCOMs, namely those reported by 
\citet{Bianchi2019a} plus C$_2$H$_5$CN, a-(CH$_2$OH)$_2$, and HCOCH$_2$OH.
Both the ASAI and the CALYPSO analysis led to an emitting size, 
for iCOMs, of 0$\farcs$3, perfectly consistent with what 
is found in the present analysis of S-bearing species (see Table 2).
The column densities are also well in agreement, being different
by less than a factor of 2 for all the iCOMs in common except for
methanol and formamide (a factor of 3--4).
Very recently, also \citet{Yang2021} reported the results
of the PEACHES ALMA survey on the chemical content of star forming regions in Perseus, SVS13-A among them. A large number of iCOMs are detected, in agreement with the ASAI and CALYPSO results.
Column densities, measured in this case of a 0$\farcs$5 source size, 
are also consistent considering the uncertainties.
In conclusion, once observed with a $\sim$ 1$\arcsec$ resolution,
it looks that S-bearing species are emitting from a region similar
to that associated with the hot corino chemistry.

In order to derive the abundances of S-bearing species 
we assumed N$_{\rm H_2}$ = 3 $\times$ 10$^{24}$ cm$^{-2}$
as the typical value for the inner 0$\farcs$3 region,
following what was adopted by \citet{Bianchi2019a} 
and \citet{Belloche2020} for the iCOMs
analysis, using the continuum images by \citet{Chen2009}.
The abundances are reported in Table 2: the most abundant S-species
are CS (0.3--6 $\times$ 10$^{-6}$), SO (7 $\times$ 10$^{-9}$ -- 1 $\times$ 10$^{-7}$),
and SO$_2$ (1--10 $\times$ 10$^{-7}$). On the other hand,
H$_2$CS and OCS have similar, lower, abundances (a few 10$^{-10}$), while X$_{\rm NS}$ $\sim$ 10$^{-10}$--10$^{-9}$.
Interestingly, the H$_2$CS abundance in SVS13-A is four orders of magnitude higher with respect to the recently measured in the
H$_2$CS ring in the protoplanetary disk around the 
Class II HL Tau \citep{Codella2020}.
This strongly supports that sulfur chemistry is indeed
definitely evolving during the star-forming process.

Finally, Figure 15 summarises the abundances derived here for the
sulfuretted molecules as well as (i) those of iCOMs reported by \citet{Bianchi2019a}, (ii) and those derived using the CALYPSO datatset \citet{Desimone2017,Belloche2020} applying the same 
H$_2$ column density. For completeness, we also added 
5-atoms molecules such as H$_2$CCO \citep{Bianchi2019a}, and CH$_3$CN, and NH$_2$CN \citep{Belloche2020}.
The present census of the S-bearing molecules 
thus contribute to building up a suite of  
abundances representative of the inner 90 au SVS13-A region,
calling for modelling (out of the scope of this paper) to constraint the chemical evolution in star-forming regions. 
Obviously, any theoretical approach in modelling the SVS13-A
chemical richness will face the long-standing problem on the
main reservoir of sulfuretted molecules on dust mantles.
A progress in that direction would surely unlock
the interpretation of the chemical richness (also) around protostars.
Meanwhile, a little step can be done
by comparing the H$_2$CS emission of H$_2$CO, which 
is not properly an iCOM, but is key molecule for 
the production of more complex species
\citep[see e.g.][and references therein]{Caselli2012,Jorgensen2020}.

\begin{figure}
\centerline{\includegraphics[angle=0,width=8.7cm]{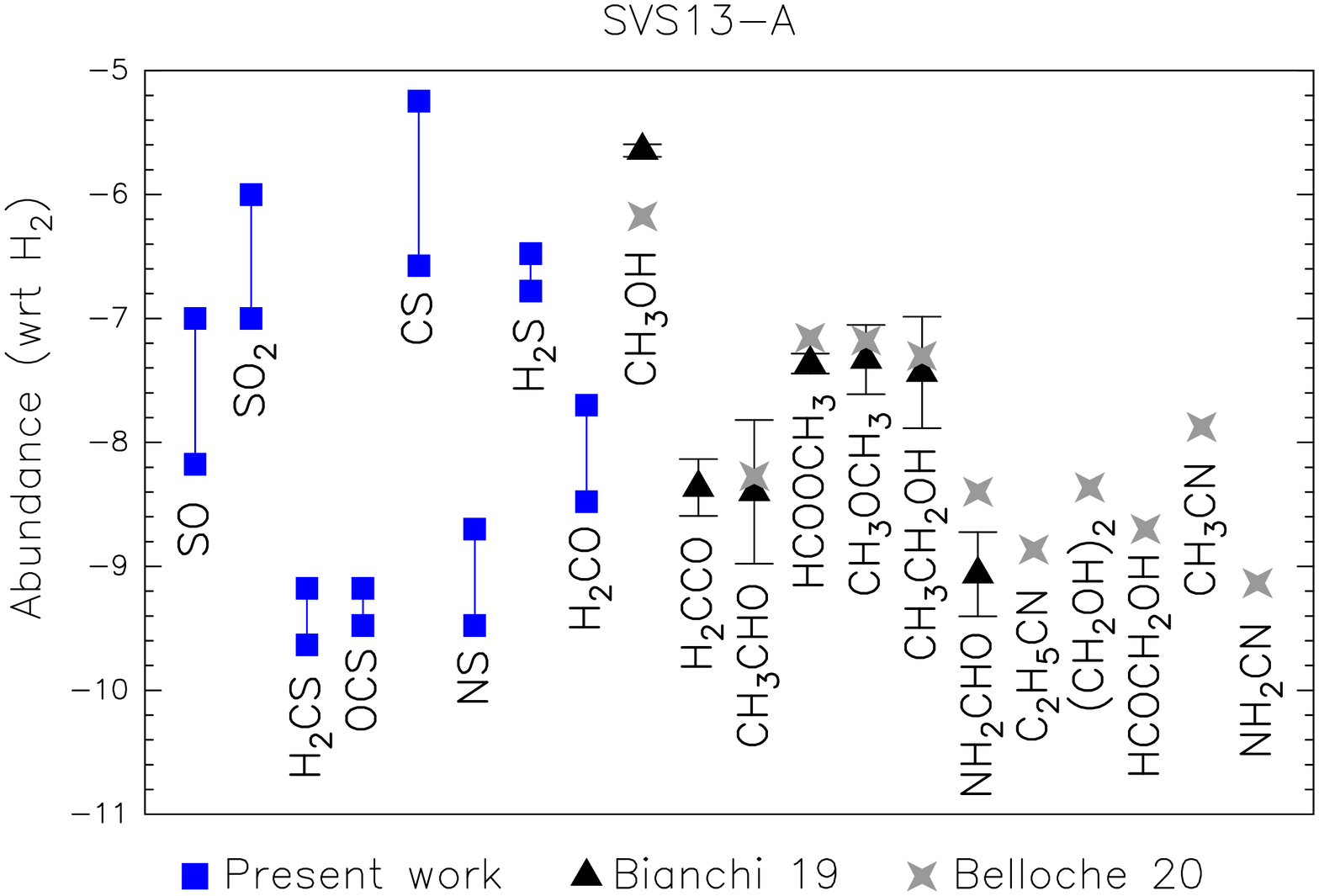}}
\caption{Summary of the abundances with respect to H$_2$
derived for the hot corino observed towards SVS13-A
\citep[N$_{\rm H_2}$ = 3 $\times$ 10$^{24}$ cm$^{-2}$][ ]{Chen2009}. 
Blue points are for the S-species reported in the present paper,
while blue triangles \citep{Bianchi2019a} and grey diamonds
\citep{Belloche2020} are for iCOMs (see Sect. 6.1).}
\end{figure}

\subsection{H$_2$CS versus H$_2$CO} 

One of the main reasons of astrochemical studies of protostellar regions is to understand the chemical composition of the gas where planets start their 
formation process. As a matter of fact, to understand
if the planetary composition has the marks  of where planets formed would definitely be a breakthrough result
\citep[e.g.][and references therein]{Turrini2021}. 
Obviously, several processes are expected to be at work to sculpt
the characteristics of planetary atmospheres. Many intermediate steps should intervene between the interstellar chemistry and the planetary chemistry, and consequently, a complete chemical reset cannot be excluded. 
Whatever is the path
leading to planets, the elemental abundance ratios are the root where the atmospheric chemistry is anchored
\citep[e.g.][and references therein]{Booth2019, Cridland2019}. 
Together with O, C, N, sulfur is playing a major role in this aspect \citep[e.g.][]{Semenov2018, Fedele2020, Turrini2021}.
With time, the number of S-bearing species around protoplanetary disks is definitely increasing.
First, several detections of CS and SO has been reported  \citep[e.g.][]{Dutrey1997,Dutrey2017,Fuente2010,Guilloteau2013,Guilloteau2016,Pacheco2016}.
More recently, thanks to the advent of ALMA, also
H$_2$S, H$_2$CS have been imaged towards disk around Class I/II
objects showing rings and gaps \citep[][]{Phuong2018,LeGal2019, Codella2020,Loomis2020}. These findings allow us to
compare what we found for the Class I SVS13-A with more
evolved star forming regions, namely Class I/II (10$^5$--10$^6$ yr old). 

In this context, as recently remarked by \citet{Fedele2020},
the abundance ratio between H$_2$CS and H$_2$CO could be used
to investigate the S/O abundance ratio provided that both molecules
in the gas-phase are mainly formed by reacting O or S with the 
methyl group CH$_3$. In order to estimate [H$_2$CS]/[H$_2$CO] in SVS13-A,
we used the emission of the o--H$_2$CO(6$_{1,5}$--6$_{\rm 1,6}$) transition at 101332.991 MHz (E$_{\rm u}$ = 88 K; from the 
Cologne Database for Molecular Spectroscopy \citep[CDMS, ][]{Muller2001,Muller2005}.
falling in the present Setup 6 at 3mm. Figure 16 reports the H$_2$CO spatial distribution as well as the spectrum extracted at the peak emission, both perfectly consistent
with that of H$_2$CS as observed with the same spatial 
and spectral resolution (see Figs. 10 and 11).
We then adopted the same size and kinetic temperature range inferred for H$_2$CS (see Table 2), the ortho/para ratio equal to 3 and, assuming LTE conditions, we derived a total column density N$_{\rm H_2CO}$ = 2-8 $\times$ 10$^{17}$ cm$^{-2}$. The [H$_2$CS]/[H$_2$CO] ratio is then between
9 $\times$ 10$^{-4}$ and
2 $\times$ 10$^{-2}$.

The present values obtained for a Class I object 
can be compared with other measurements obtained from interferometric observations sampling
Solar System scales around protostars at different evolutionary stages:

\begin{itemize}

\item the prototypical Class 0 hot corino 
IRAS 16293-2422B as revealed with ALMA on a 60 au scale
\citep[PILS project:][]{Jorgensen2016,Persson2018,Drozdovskaya2018,Drozdovskaya2019}:
7 $\times$ 10$^{-4}$;

\item  late Class I or II protoplanetary disks as imaged
with ALMA on a 40 au scale \citep[ALMA-DOT project:][]{Podio2020-iras,Codella2020,Garufi2021}: 0.1--0.2 (HL Tau), and 0.4--0.7 (IRAS 04302+2247).

\end{itemize}

The comparison suggests an increase with time of the
[H$_2$CS]/[H$_2$CO] ratio in the gaseous compositions around stars by more than one order 
of magnitude. Obviously, we cannot quantify [S]/[O] from [H$_2$CS]/[H$_2$CO] given the complexity of the 
overall S and O chemistry, but the present 
findings suggest
that [S]/[O] could change along the Sun-like star forming process. 

Finally, we inspected
what has been measured with the ROSINA spectrometer towards the comet 
67P/Churyumov-Garasimenko (C-G) in the context of the ESA {\it Rosetta} space mission \cite{Rubin2020}.
ROSINA derived the chemical composition of the volatiles in the coma, reporting the H$_2$CS and H$_2$CO abundance with respect to water. 
The corresponding [H$_2$CS]/[H$_2$CO] ratio ranges
in the 7 $\times$ 10$^{-4}$ -- 4 $\times$ 10$^{-2}$
range, i.e. a quite wide spread which does not allow us to
verify if a relic of the early stages of our
Solar System supports the tentative [H$_2$CS]/[H$_2$CO]
dependence on time from Class 0 to Class II objects.
Clearly, more measurements are needed to perform a  statistical study.

\begin{figure}
\centerline{\includegraphics[angle=0,width=7cm]{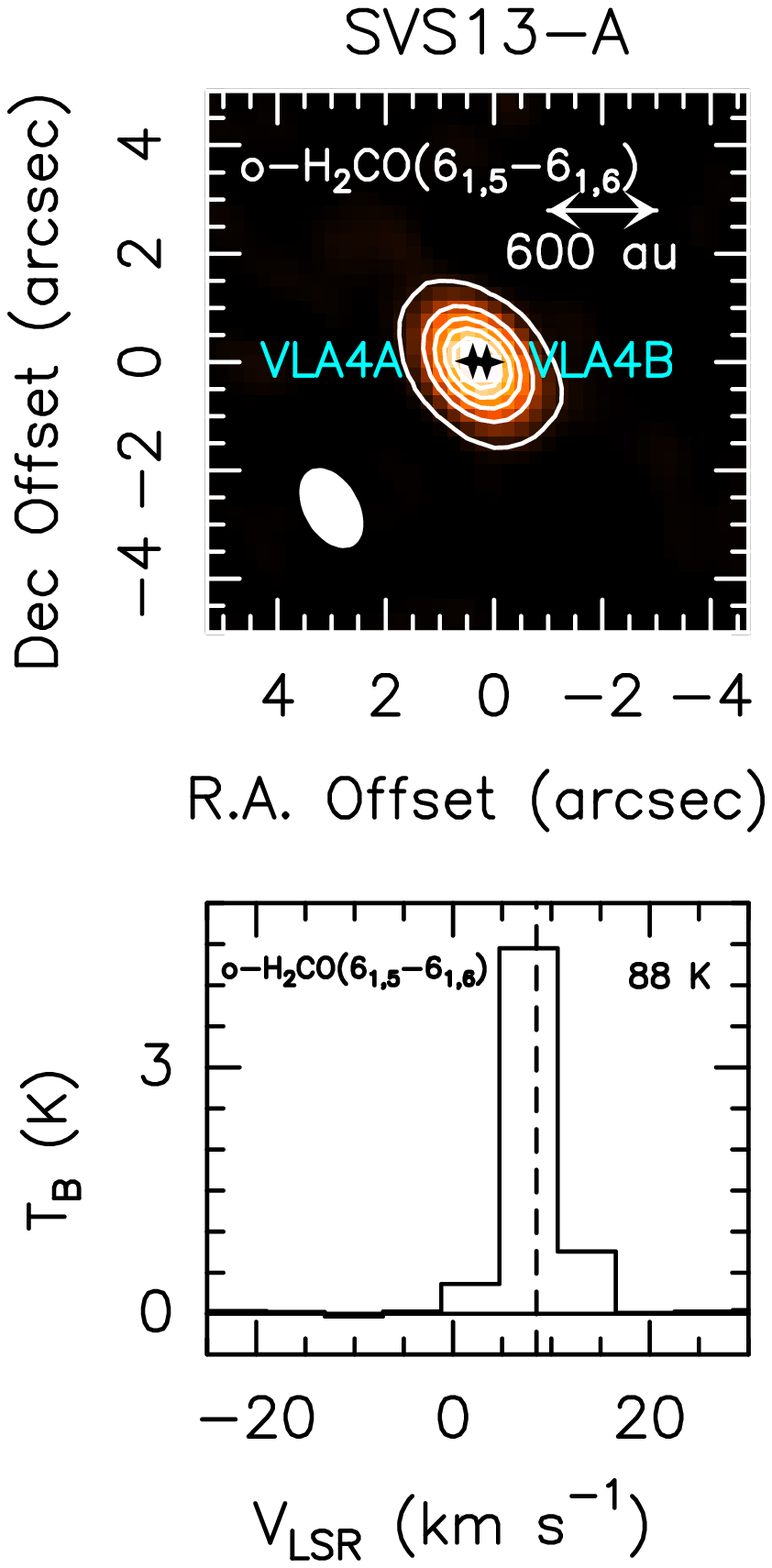}}
\caption{{\it Upper panel:} Spatial distribution of the
SVS13-A region obtained using the o--H$_2$CO(6$_{\rm 1,5}$--6$_{\rm 1,6}$)  emission. 
The whole velocity emitting region has been used (from --5 km s$^{-1}$
to +20 km s$^{-1}$).
The angular offsets are relative to the phase center
(see Sect. 2).
First contour and steps are, respectively, 5$\sigma$
(25 mJy km s$^{-1}$), and 20$\sigma$, respectively.
The filled ellipse shows the synthesised beam (HPBW): 1$\farcs$6 $\times$ 1$\farcs$1 (PA= 41$^\circ$).
The black crosses indicate the positions of the VLA4A and VLA4B 
sources as imaged using the VLA array by \citet{Tobin2018}.
{\it Lower panel:} Observed o--H$_2$CO(6$_{\rm 1,5}$--6$_{\rm 1,6}$) spectrum 
(in T$_{\rm B}$ scale, see Table 1) extracted at the emission peak: $\alpha_{2000}$ = 03$^h$29$^m$03$^s$.75,  $\delta_{2000}$ = +31$^\circ$16$'$03$''$.8. 
The upper level energy is reported.
The vertical dashed line stands for the ambient LSR velocity (+8.6 km s$^{-1}$, \citet{Chen2009}).}
\end{figure}

\subsection{NS versus NS$^+$}

The cation NS$^+$ has been very recently discovered in interstellar space
using IRAM 30-m observations in the mm-spectral window of a sample of low-mass star-forming regions \citep{Cernicharo2018}. More specifically,  NS$^+$ has been revealed
towards cold molecular clouds and prestellar cores as well as in shocked protostellar 
regions and in the direction of hot corinos (e.g. NGC1333-IRAS4A).
\citet{Cernicharo2018} derived the NS/NS$^+$ ratio, lying in the 30--50 range,
which has been modelled adopting the following chemical routes
\citep[see e.g. ][]{agundez2013}:
NS is formed by reaction of SH with N or of NH with S, while NS$^+$ is formed when the atomic nitrogen reacts with the SO$^+$ and SH$^+$ ions. 
On the other hand, both NS and NS$^+$ should be destroyed via reactions 
with the O, N, or C atoms. NS$^+$ should go first through dissociative recombination.
\citet{Cernicharo2018} fit the observations by using a relatively cold gas, even 
for the NGC1333-IRAS4A region (T$_{\rm kin}$ $\simeq$ 30 K), which in turn means that NS$^+$ is not released in the hot corino, but instead from a more extended  molecular envelope. 

In the present SOLIS observations of SVS13-A, NS is clearly tracing the protostellar region with a source size less than 100 au, then it is plausibly emitted in the hot corino (see Fig. 12). 
In fact, in Sect. 4 we derived the NS column density by assuming a kinetic temperature in the 50--300 K range.

Interestingly, the present SOLIS NOEMA setup is covering (with
a $\sim$ 1$\farcs$3 spatial resolution) the frequency of the NS$^+$(2--1) line (100198.55 GHz), characterised by E$_{\rm u}$ = 7 K, $S\mu^2$ = 9 D$^2$ \citep{Cernicharo2018}. However, this line was not detected and only an upper limit on NS$^+$ column density can be derived by assuming
the same emitting size as well as the same kinetic temperatures adopted for the NS analysis (see Table 2). The 3$\sigma$ upper limit on the 
velocity integrated emission is 1 K km s$^{-1}$, which in turn allows
us to fix an upper limit on N$_{\rm NH^+}$ $\leq$ 8 $\times$ 10$^{14}$ cm$^{-2}$.
In summary, we can derive N(NS)/N(NS$^+$) $\geq$ 10, a number 
in agreement with those reported by 
\citet{Cernicharo2018}, in particular for the NGC1333-IRAS4A hot corino ($\sim$ 40).
Note that these findings are also consistent with the inspection of the 
spectra of the IRAM 30-m ASAI Legacy \citep{Lefloch2018}, which provides an unbiased spectral survey at 1, 2, and 3mm of SVS13-A.
ASAI covered not only the NS$^+$(2-1) line, but also the 
$J$ = 3--2 and 5--4 ones, at 150295.607 MHz and 250481.463 MHz, respectively
\citep{Cernicharo2018}. No detection has been found, providing 
an upper limit on the NS$^+$ column density of a few 10$^{14}$ cm$^{-2}$,
similarly to what was obtained with NOEMA SOLIS.
The present findings, providing the first constraint on the
N(NS)/N(NS$^+$) ratio based on interferometric data, call for a comparison with predictions from astrochemical modelling at work at kinetic 
temperatures typical of hot corinos ($\geq$ 100 K). 


\subsection{SO versus SO$_2$} 

The importance of the SO$_2$ over SO abundance ratio in star forming regions
has been discussed since the end of the last century.
The starting assumption was that H$_2$S is the main S-bearing species
in the dust mantles. As a consequence, the injection of the material frozen on ices due either to thermal heating (hot cores, hot corinos)
or due to sputtering (shocks), or even for
chemical desorption caused by the excess energy
of an exothermic reaction \citep{Oba2018}, increases the H$_2$S abundance in the gas-phase, which in turn forms first SO and successively SO$_2$
\citep[e.g.][]{Pineau1993,Charnley1997,Hatchell1998}.
As a result, [SO$_2$]/[SO] has been proposed to be a chemical clock
to date the evaporation/sputtering process. 
However, successive attempts to apply this tool in young 
($\geq$ 10$^3$ yr) shocked regions did not result to be efficient 
\citep{Codella1999,Wakelam2004,Codella2005} given all the
uncertainties associated with the S-chemistry starting from 
the still open question on the main S-bearing species frozen on ices
\citep[e.g.][and references therein]{Laas2019,Taquet2020}.
On the other hand, a Kitt Peak single-dish (HPBW = 43$\arcsec$) 
survey of Class 0 and Class I hot corinos (10$^4$--10$^5$ yr) by \citet{Buckle2003} suggested an evolutionary trend, with [SO$_2$]/[SO] $\simeq$ 0.1 for Class 0 and [SO$_2$]/[SO] $\simeq$ 0.4 for Class I. 

Moving to interferometric observations, the present dataset shows
for the Class I SVS13-A target [SO$_2$]/[SO] $\sim$ 10, a value 
larger than what measured with ALMA towards the Class 0 IRAS16293-2422 hot corino \citep[PILS:][]{Drozdovskaya2018,Drozdovskaya2019}:
[SO$_2$]/[SO] ratios $\sim$ 3. Both measurements 
are higher than what measured using IRAM-NOEMA (in the SOLIS context) of the young shocked regions  L1157 and L1448 outflows: [SO$_2$]/[SO] $\simeq$ 0.2 \citep[NGC1333-IRAS4A][]{Taquet2020}, and 
0.1--0.3 \citep[L1157][]{Feng2020}.
In conclusion, these recent findings 
support the use  
the [SO$_2$]/[SO] ratio to date gas enriched in sulphur around protostars, 
using 
high-spatial interferometric images to disentangle contributions
due to different physical components.

\subsection{On the sulfur budget in SVS13-A}

The recent results by \citet{Kama2019} and \citet{Shing2020} 
indicates that a large fraction ($\sim$ 90\%) of sulphur in dense star forming regions is in very refractory forms such as FeS
or S$_{\rm 8}$. In this context, in order to evaluate the sulfur budget in the Class I SVS13-A hot corino,
we need an estimate of the abundance of H$_2$S, which, as reported in the Sect. 1, is postulated to be a major S-bearing molecule on dust mantles. 
Unfortunately, the NOEMA SOLIS spectral windows do not cover
H$_2$S line frequencies. However, the
H$_2$S column density can be measured using the line 
of the para 2$_{2,0}$--2$_{1,1}$ transition observed 
at 1.4mm in the context of the IRAM 30-m ASAI Large Program \citep{Lefloch2018}.
The line is emitting at 216710.4365 MHz: the HPBW is 11$\arcsec$
and the energy of the upper level is quite high, 84 K
\footnote{The Frequencies and spectroscopic parameters have been provided by \citet{Cupp1968}, \citet{Burenin1985}, and \citet{Belov1995} and retrieved from the Cologne Database for Molecular Spectroscopy \citep{Muller2001,Muller2005}}.
This combination minimises the contaminations expected from
the cold envelope with respect to the SVS13-A hot corino, 
and excludes any possible contribution due to SVS13-B (see Fig. 1). 
Figure 17 reports the observed spectra, with the 
p--H$_2$S(2$_{2,0}$--2$_{1,1}$) profile peaking at +8.1(0.1) km s$^{-1}$, with a FWHM line width equal to 3.2(0.1) km s$^{-1}$,
and an integrated area (in T$_{\rm MB}$ scale) of 534(14) mK km s$^{-1}$.
Given S$\mu^2$ = 2.1 D$^2$, assuming LTE conditions, an emitting size equal to 0$\farcs$3 and a temperature range of 50--300 K (as for the other S-species here imaged by NOEMA), 
an ortho/para ratio equal to 3, we obtain 
a total column density of 0.5--1 $\times$ 10$^{18}$ cm$^{-2}$.
Assuming again N$_{\rm H_2}$ = 3 $\times$ 10$^{24}$ cm$^{-2}$
\citep{Chen2009}, the
H$_2$S abundance results to be 2--4 $\times$ 10$^{-7}$.
Table 2 and Fig. 15 summarises the column density and abundances of the S-bearing molecules observed in the SVS13-A hot corino. Overall, if we take all of them into account we reach
[S]/[H] = 3.0 $\times$ 10$^{-7}$ -- 3.8 $\times$ 10$^{-6}$, i.e.
2\%--17\% of the Solar System [S]/[H] value
\citep[1.8 $\times$ 10$^{-5}$;][]{Anders1989}.
Obviously, not all the S-species observed around
low-mass protostars have been revealed in the present survey, missing mainly CCS as well as the SO$^+$ and HCS$^+$ ions. However,
we can reasonably assume that the contribution due to 
CCS (a standard envelope tracer), and the SO$^+$ and HCS$^+$ ions (two orders of magnitude less abundant than CS in the warm
shocked gas in the L1157 outflow \citet{Podio2014}) do not
significantly contribute to the total S budget in SVS13-A. 

The present measurements can be compared with what obtained in surveys of other prototypical regions associated with Sun-like star-forming regions:

\begin{itemize}

\item
the Taurus dark cloud TMC 1 has been recently investigated 
in the context of the GEMS project 
\citep{Fuente2019} using the IRAM 30-m and Yebes 40-m antennas in CS, SO, and HCS$^+$ (plus rarer isotoplogues of these molecules).
The authors found a strong sulfur depletion both in the traslucent low-density phase (where [S]/[H] is 2\%--12\% of the Solar System
one, very similar to what is found in SVS13-A) and in the dense core (where this percentage falls down to 0.4\%). Additional 
GEMS H$_2$S observations reported by \citet{Navarro2020} did not significantly change these results;

\item
the L1689N star forming region, hosting the well known
Class 0 hot corinos IRAS16293-2422A and B has been sampled using SO, and SO$_2$, and H$_2$S as observed 
by \citet{Wakelam2004a} with the IRAM 30-m single-dish.
The authors estimate, for the hot corino region, 
the [S]/[H] ratio to be $\sim$ 8\% of the Solar System value. 
Successively, in the context of the PILS  Large Program
\citep{Jorgensen2016}, \citet{Drozdovskaya2018,Drozdovskaya2019} report the census of S-bearing species as observed with ALMA towards the hot corino IRAS16293-2422B (SO, SO$_2$, H$_2$S, CS, H$_2$CS, 
and CH$_3$SH). Assuming an H$_2$ column density of 10$^{25}$ cm$^{-2}$ derived in the PILS context by \citep{Jorgensen2016}, the [S]/[H] ratio is then $\sim$ 1\% of the Solar System measurement;

\item 
\citet{Holdship2019} measured the sulfur budget in the $\sim$ 100 K
shocked regions associated with the protostellar L1157 jet.
In this case, the sampled region is chemically enriched due to
sputtering and shuttering induced by a shock with a velocity
of $\simeq$ 20--40 km s$^{-1}$ \citep[e.g.][]{Flower2010,Viti2011}.
For the bright L1157-B1 shock \citet{Holdship2019} report
[S]/[H] $\sim$ 10\% of the Solar System, a value in agreement with
what found in SVS13-A.

\end{itemize} 

\begin{figure}
\centerline{\includegraphics[angle=0,width=8.7cm]{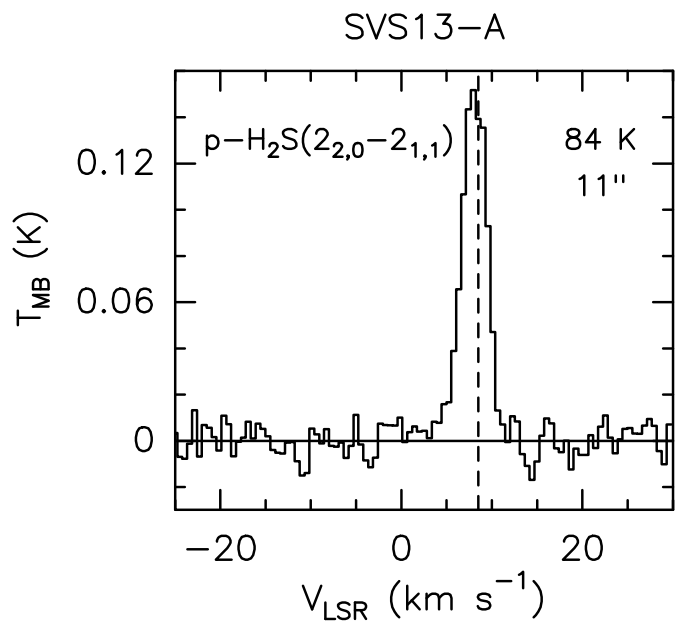}}
\caption{Spectrum of the p--H$_2$S(2$_{2,0}$--2$_{1,1}$) line
(in T$_{\rm MB}$ scale) as observed towards SVS13-A at 1.4mm using the IRAM 30-m antenna.
Transition and corresponding upper level energy are reported. The HPBW is also indicated. The vertical dashed line stands for the ambient LSR velocity (+8.6 km s$^{-1}$, \citet{Chen2009}).}
\end{figure}

In summary, the measurements of percentage of the [S]/[H] value with 
respect to the Solar System value are: 0.4\% (dense core),
1\%-8\% (Class 0 object), 2\%-17\% (Class I object),
10\% (shock). 
Obviously, these estimates are strongly dependent on the
uncertainties associated with the adopted H$_2$ column densities.
For instance \citet{Jorgensen2016} use N(H$_2$) = 10$^{25}$ cm$^{-2}$ for
IRAS16293-2422B, while \citet{Bianchi2019a} and \citet{Belloche2020} adopt
3 $\times$ 10$^{24}$ cm$^{-2}$ for SVS13-A. 
With these caveats in mind, we can speculate that
(i) the increase of S-bearing molecules in the gas-phase
due to thermally evaporation in a Class 0 hot corino seems to keep
constant also in the more evolved Class I stage; 
(ii) the enrichment of sulfuretted molecules (with respect to
dense cold clouds) due to thermal evaporation seems to be
quantitatively the same than that in shocks, where sputtering rules.

\section{Summary and conclusions}\label{sec:conclusions}

In  the  context  of  the  IRAM  NOEMA  SOLIS  Large  Program 
we observed the SVS13-A Class I object at both 3mm (beam $\simeq$ 1$\farcs$5) and 1.4mm
(beam $\simeq$ 0$\farcs$6) in order to obtain a census of the S-bearing species. The main results can be summarised as follows:

\begin{itemize}

\item 
We obtained 32 images of emission lines of $^{32}$SO, $^{34}$SO, C$^{32}$S, C$^{34}$S, C$^{33}$S, OCS,  H$_2$C$^{32}$S, H$_2$C$^{34}$S, and NS, sampling $E_{\rm u}$ up to 244 K.
The low-excitation (9 K) SO is peaking towards
SVS13-A. This emission traces also
the molecular envelope and the low-velocity outflow emission driven by SVS13-A. 
SO is also imaging the collimated high-velocity (up to 100 km s$^{-1}$ shifted with respect to the systemic velocity) jet driven by the nearby SVS13-B Class 0. 
In general, the molecular envelope contributes to 
the low-excitation (less than 40 K) SO and CS emission.
Conversely, all the rest of the observed species and transitions show
a compact emission around the SVS13-A coordinates indicating
the inner $\sim$ 100 au protostellar region.

\item
The non-LTE LVG analysis of SO, SO$_2$, and
H$_2$CS indicates a hot corino origin.
The emitting size is about 90 au (SO),
60 au (SO$_2$), and 120 au (H$_2$CS).
For SO, we have
$T_{\rm kin}$ $\geq$ 150 K, and $n_{\rm H_2}$ $\geq$ 6 $\times$ 10$^{6}$ cm$^{-3}$. 
The analysis of SO$_2$ leads 
$T_{\rm kin}$ $\simeq$ 100--300 K, 
and $n_{\rm H_2}$ $\geq$ 5 $\times$ 10$^{6}$ cm$^{-3}$. 
Finally, for  H$_2$CS, $T_{\rm kin}$ $\geq$ 50 K, 
and $n_{\rm H_2}$ $\geq$ 10$^{5}$ cm$^{-3}$. 
For OCS we used the rotation diagram approach
using LTE obtaining a $T_{\rm rot}$ of 120$\pm$50 K,
consistent with the temperatures expected for a 
hot corino.

\item 
The bright emission from S-bearing molecules  is confirmed to arise from the SVS13-A hot corino, where the iCOMs, previously imaged, are abundant \citet{Desimone2017,Bianchi2019a, Belloche2020}. 
The abundances of the sulphuretted species are in the following ranges:
0.3--6 $\times$ 10$^{-6}$ (CS), 
0.7--10 $\times$ 10$^{-7}$ (SO),
1-- 10 $\times$ 10$^{-7}$ (SO$_2$), a few 10$^{-10}$ 
(H$_2$CS and OCS), and 10$^{-10}$--10$^{-9}$ (NS).

\item Also NS is tracing a region with a size less than 100 au. We constrain for the first time the N(NS)/N(NS$^+$) 
using interferometric observations: $\geq$ 10. 
This is in agreement with what is previously reported for
the NGC1333-IRAS4A hot corino by 
\citet{Cernicharo2018} using single-dish measurements, supporting that 
NS$^{+}$ is mainly formed in the extended envelope.

\item Once measured using the NOEMA array, the [SO$_2$]/[SO] ratio towards SVS13-A is $\sim$ 10, a value 
slightly larger (by a factor 3) than what measured with ALMA towards the Class 0 IRAS16293-2422B hot corino \citep[PILS:][]{Drozdovskaya2018,Drozdovskaya2019}.
This is clearly not enough to support the
use of the [SO$_2$]/[SO] as chemical clock.
However, after several unsuccessful attempts done 
in the past using
observations sampling gas on large scales (mainly using single-dish antennas), the present measurements
suggest to verify the use of [SO$_2$]/[SO] as chemical clock, but sampling the inner 100 au
around the protostars.

\item
Considering that in the gas-phase H$_2$CS and H$_2$CO (i)
are mainly formed by reacting O or S with the 
methyl group CH$_3$, and (ii) they are among the fews 
species detected from protostars to protoplanetary disks, we derived their abundance ratio
in SVS13-A, obtaining 0.9 $\times$ 10$^{-3}$ -- 2 $\times$ 10$^{-2}$.
The comparison between the few interferometric observations sampling Solar System scales (IRAS 16293-2422B, SVS13-A,
HL Tau, and IRAS 04302+2247)
suggests an increase with time of the
[H$_2$CS]/[H$_2$CO] ratio in the gaseous compositions around stars by more than one order
of magnitude.
It is then tempting to speculate
that [S]/[O] could vary along the Sun-like star forming process.

\item
The estimate of the [S]/[H] budget in SVS13-A is
2\%-17\% of the Solar System value
(1.8 $\times$ 10$^{-5}$). This number is consistent with
what was previously measured towards Class 0 objects 
(1\%-8\%). As a matter of fact, it seems that the
enrichment of the S-bearing molecules in 
Class 0 hot corinos (with respect to
dense and cold clouds) remain in the Class I stage.

\end{itemize}

To conclude, the present results are in agreement with the results reported by \citet{Kama2019},
who analysed a sample of young stars photospheres concluding that 
89\%$\pm$8\% of elemental sulfur in their disks is locked in refractory form.
\citet{Kama2019} conclude that the main S-carrier in the dust has to be more refractory than water, suggesting sulfide minerals such as FeS.
Note also that recently \citep{Shing2020} have shown that radiation chemistry 
converts a large fraction of S in allotropic form, in particular S$_{\rm 8}$, which cannot be detected (except for possible desorption products due to photo-processes or shocks, 
such as S$_{\rm 2}$, S$_{\rm 3}$ and S$_{\rm 4}$, as also detected 
in the coma of the 67P/Churyumov-Gerasimenko comet \citep{Calmonte2016}. 
Indeed, this is supported by the fact that even in a 
shocked region such as L1157-B1, 
associated with both C and J-shocks \citep{Benedettini2012} and 
indeed rich in water \citep[H$_2$O/H$_2$ = 10$^{-4}$;][]{Busquet2014}, the 
[S]/[H] ratio is not significantly larger than what is found in both Class 0 
and Class I hot corinos.




\begin{acknowledgements}
We thank the referee M. Drozdovskaya for her careful and instructive report, that definitely improved the manuscript.
We are also very grateful to all the IRAM staff, whose dedication 
allowed us to carry out the SOLIS project. 
This work was supported 
by (i) the European MARIE SK$\L$ODOWSKA-CURIE ACTIONS under the European Union's Horizon 2020 research and innovation programme, for the Project “Astro-Chemistry Origins” (ACO), Grant No 811312, (ii) the European Research Council (ERC) under the European Union's Horizon 2020 research and innovation programme, for
the Project "The Dawn of Organic Chemistry" (DOC), grant agreement No
741002, and (iii)  the project PRIN-INAF 2016 
The Cradle of Life - GENESIS-SKA (General Conditions in 
Early Planetary Systems for the rise of life with SKA).
This work is also supported by the French National Research Agency in the framework of the Investissements d’Avenir program (ANR-15-IDEX-02), through the funding of the "Origin of Life" project of the Univ. Grenoble-Alpes.
\end{acknowledgements}

\bibliographystyle{aa} 
\bibliography{Mybib.bib} 

\clearpage 

\appendix

\section{Comparison between NOEMA-SOLIS and IRAM 30m-ASAI}

Figures A.1 and A.2 show the comparison between the IRAM 30-m 1.4mm and 3mm spectrum obtained in the context of the ASAI Large Program \citet{Lefloch2018} and the spectra derived by integrating the emission
in the NOEMA-SOLIS images in a region equal to the HPBW of the IRAM 30-m.
The output is clearly different, with the single-dish observations
collecting not only the emission imaged with NOEMA (LAS $\simeq$ 8$\arcsec$--9$\arcsec$), but also the large scale structure (e.g. molecular envelope around SVS13-A, extended outflow.). 
More specifically, Figs. A.3 and A.4 report zoomed-in portions of the
3mm and 1.4mm spectra to enlight lines of selected S-bearing molecules.
The comparison confirms that, as expected,
the NOEMA-SOLIS maps are well suited to image hot corinos. 
A particular case is represented by the SO single-dish spectra, 
which are collecting a considrable amount of flux due to extnded emission. On the other hand, the lines due to e.g. 
SO$_2$ and H$_2$CS are less affected by emission filtering.

\begin{figure*}
\centerline{\includegraphics[angle=0,width=14cm]{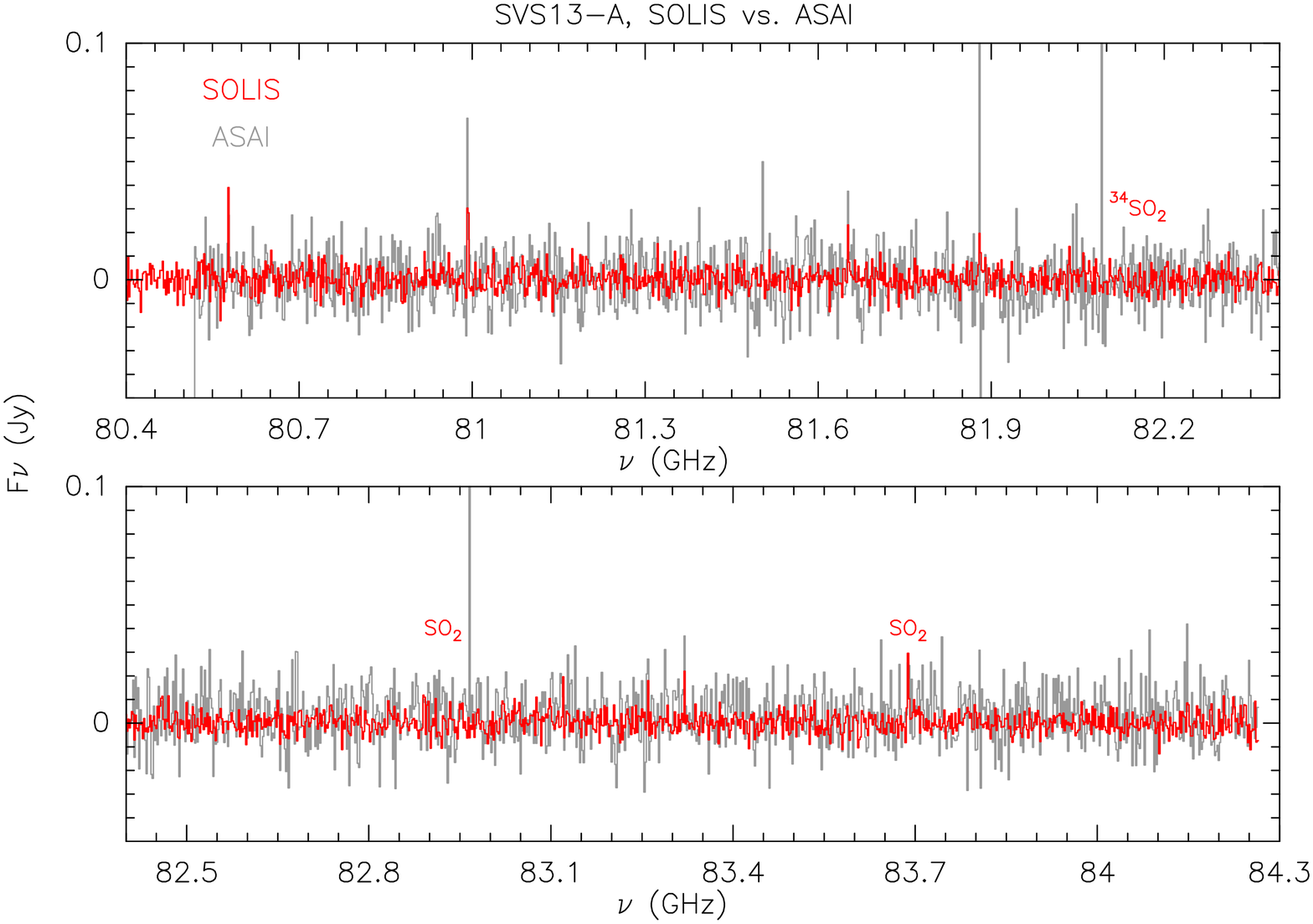}}
\centerline{\includegraphics[angle=0,width=14cm]{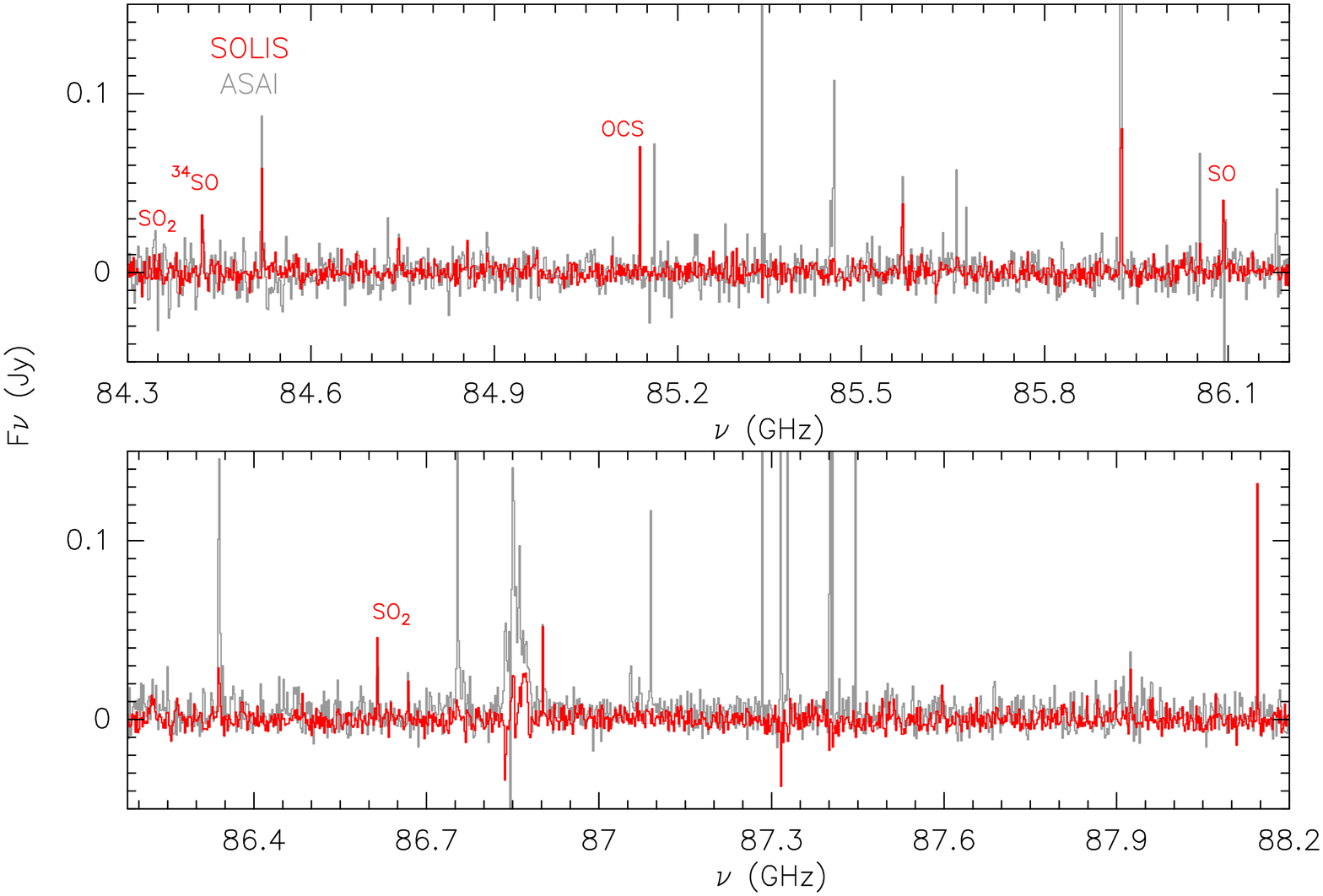}}
\caption{Comparison (in flux density scale) of the 3mm spectrum as observed using the IRAM 30-m antenna (grey; in the ASAI LP context \citet{Lefloch2018}) and that extracted from the present NOEMA SOLIS maps (red) from a circular region equal to the IRAM 30-m HPBW (24$\arcsec$ at 104 GHz, 31$\arcsec$
at 80 GHz). The ASAI spectrum has been smoothed to match the SOLIS velocity resolution (2 MHz, see Sect. 2). Red labels indicate the S-bearing species analysed in this paper (see Table 1).}
\end{figure*}

\addtocounter{figure}{-1}
\begin{figure*}
\centerline{\includegraphics[angle=0,width=14cm]{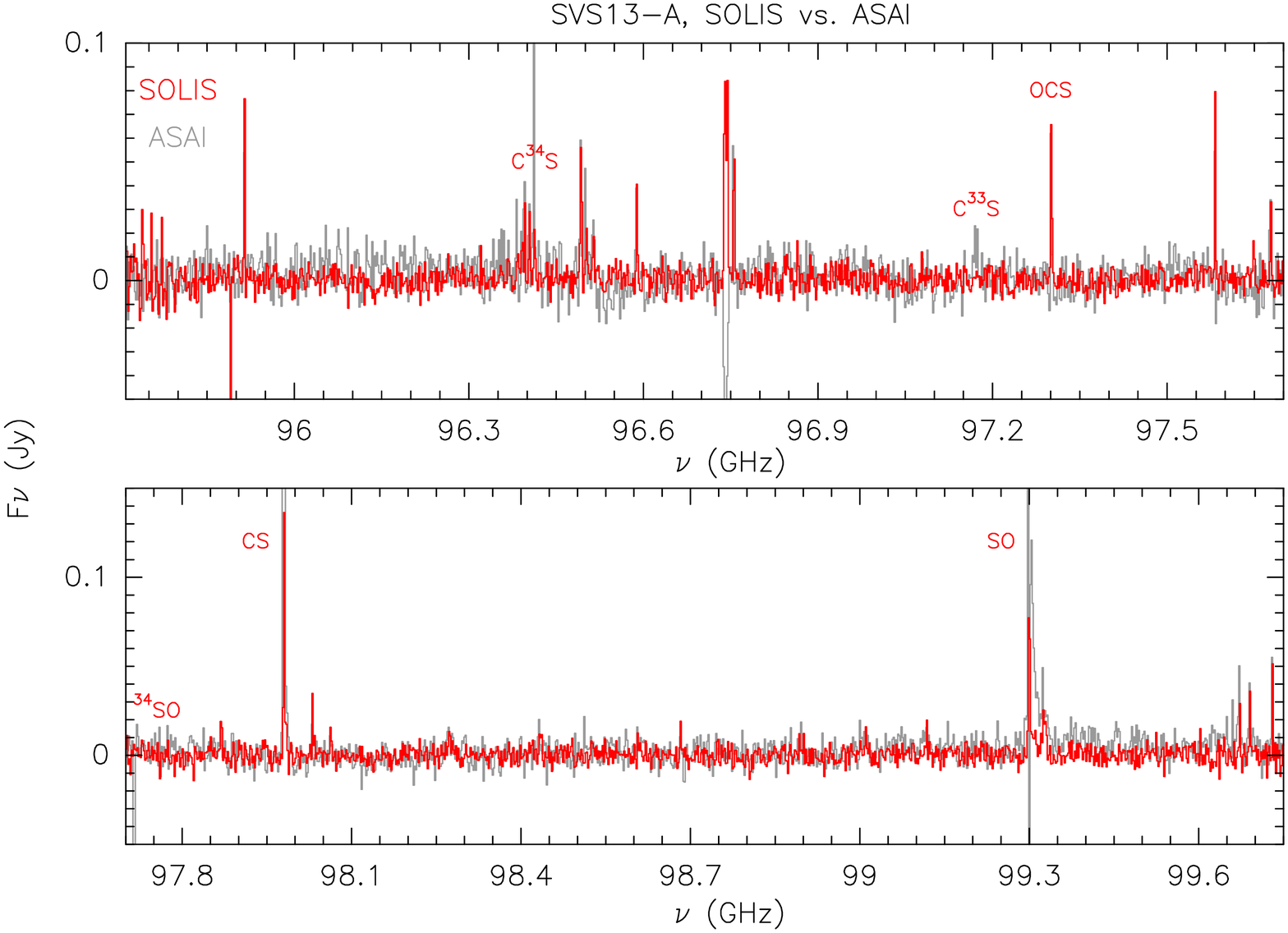}}
\centerline{\includegraphics[angle=0,width=14.5cm]{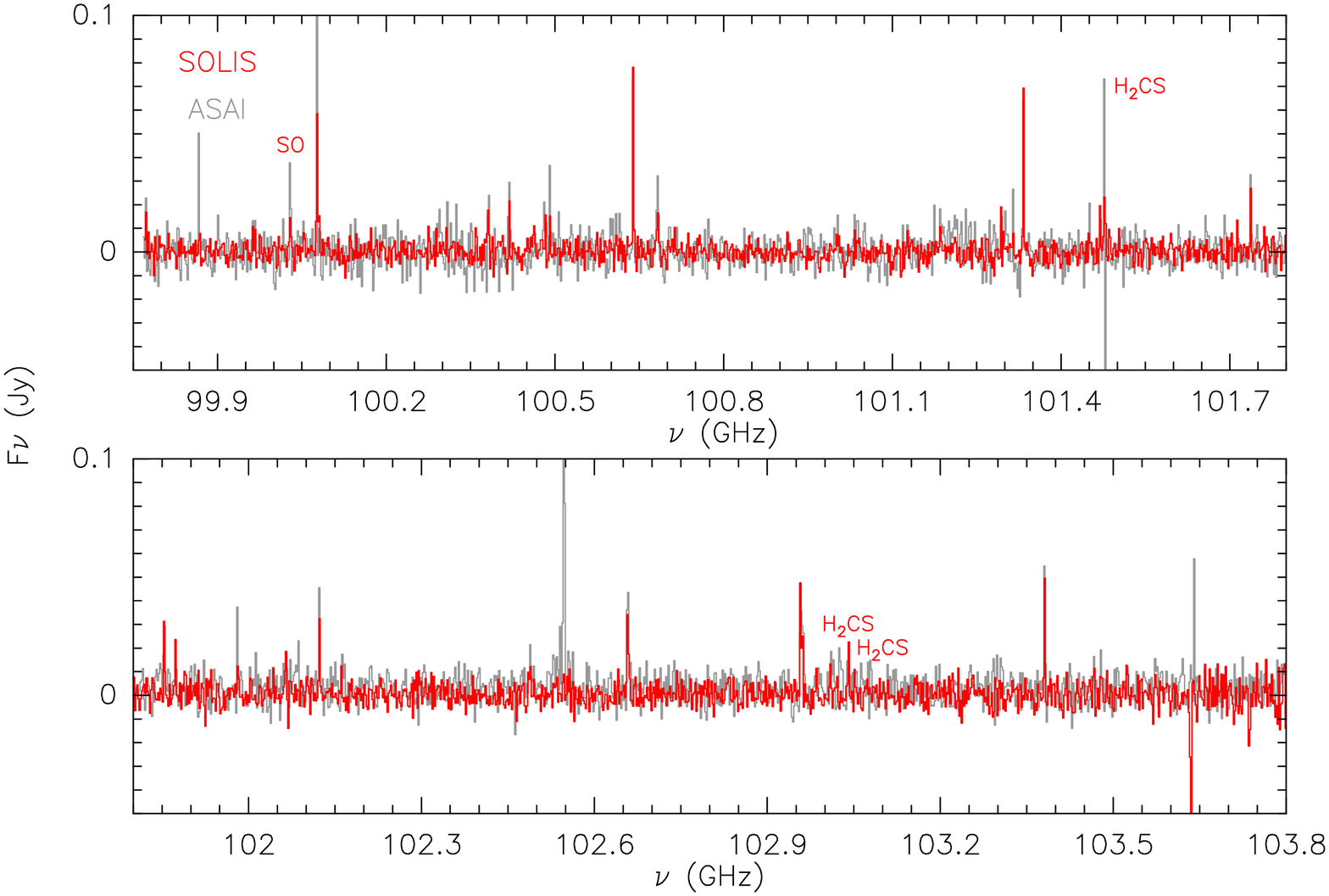}}
\caption{{\it Continued.} Comparison (in flux density scale) of the 3mm spectrum as observed using the IRAM 30-m antenna (grey; in the ASAI LP context \citet{Lefloch2018}) and that extracted from the present NOEMA SOLIS maps (red) from a circular region equal to the IRAM 30-m HPBW (24$\arcsec$ at 104 GHz, 31$\arcsec$ at 80 GHz). The ASAI spectrum has been smoothed to match the SOLIS velocity resolution (2 MHz, see Sect. 3). Red labels indicate the S-bearing species analysed in this paper (see Table 1).}
\end{figure*}

\begin{figure*}
\centerline{\includegraphics[angle=0,width=14cm]{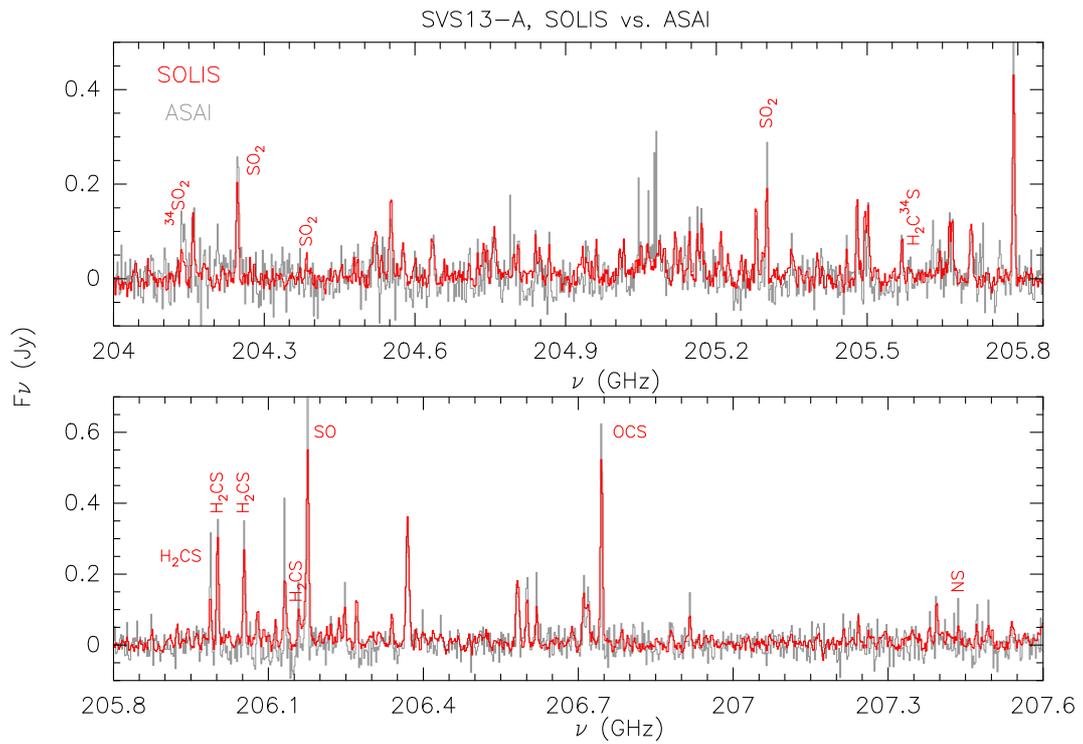}}
\caption{Comparison (in flux density scale) of the 1.4mm spectrum as observed using the IRAM 30-m antenna (grey; in the ASAI LP context \citet{Lefloch2018}) and that extracted from the present NOEMA SOLIS maps (red) from a circular region equal to the IRAM 30-m HPBW (12$\arcsec$). The ASAI spectrum has been smoothed to match the SOLIS velocity resolution (2 MHz, see Sect. 3). Red labels indicate the S-bearing species analysed in this paper (see Table 1).}
\end{figure*}

\begin{figure}
\centerline{\includegraphics[angle=0,width=9cm]{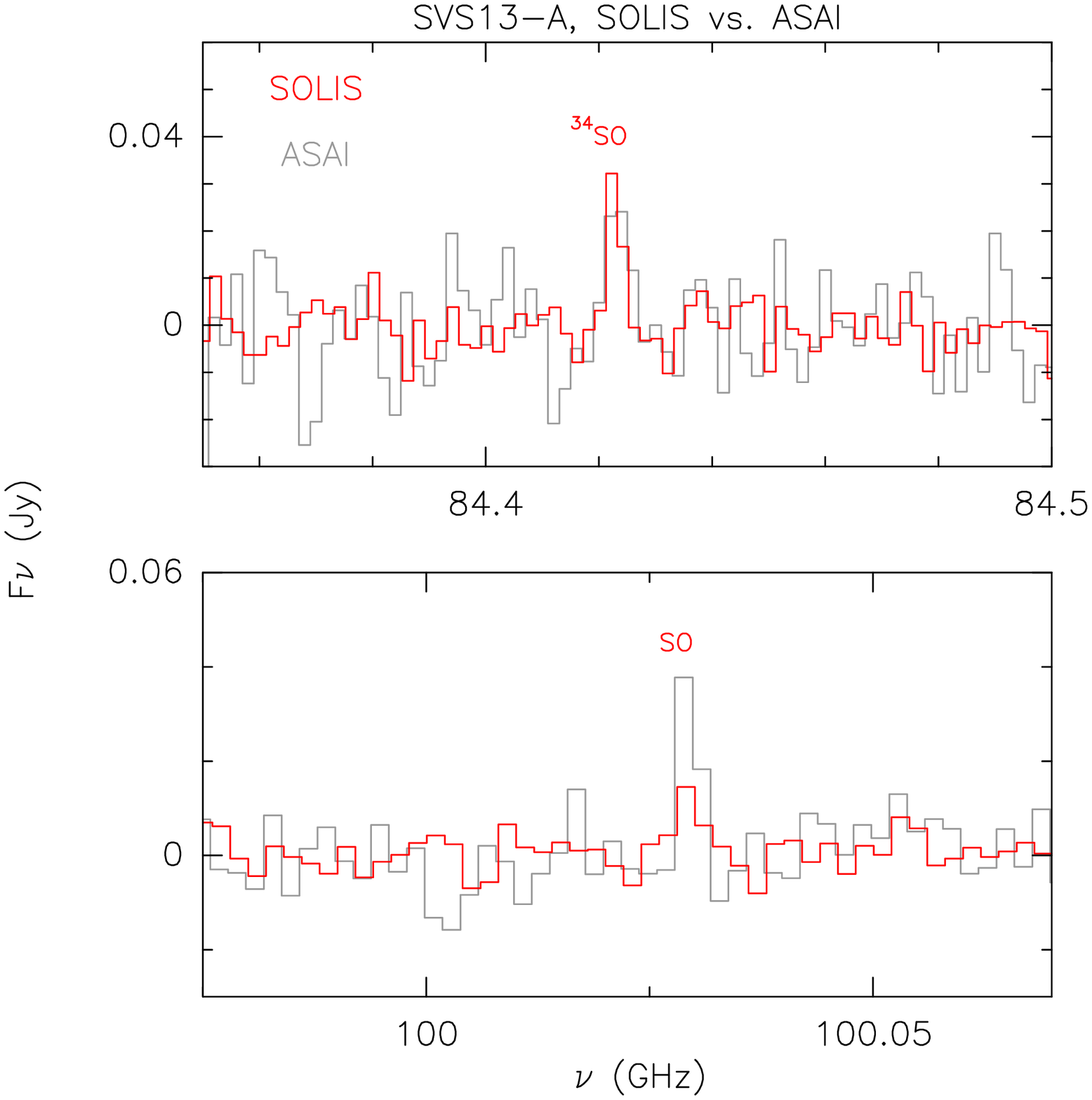}}
\caption{Zoom-in of Fig. A.1 to enlight selected emission lines of S-bearing species: comparison (in flux density scale) of the 3mm spectrum as observed using the IRAM 30-m antenna (grey; in the ASAI LP context \citet{Lefloch2018}) and that extracted from the present NOEMA SOLIS maps (red) from a circular region equal to the IRAM 30-m HPBW (25$\arcsec$ at 100 GHz, 29$\arcsec$ at 84 GHz). The ASAI spectrum has been smoothed to match the SOLIS velocity resolution (2 MHz, see Sect. 3). Red labels indicate the S-bearing  species analysed in this paper (see Table 1).}
\end{figure}

\begin{figure}
\centerline{\includegraphics[angle=0,width=9cm]{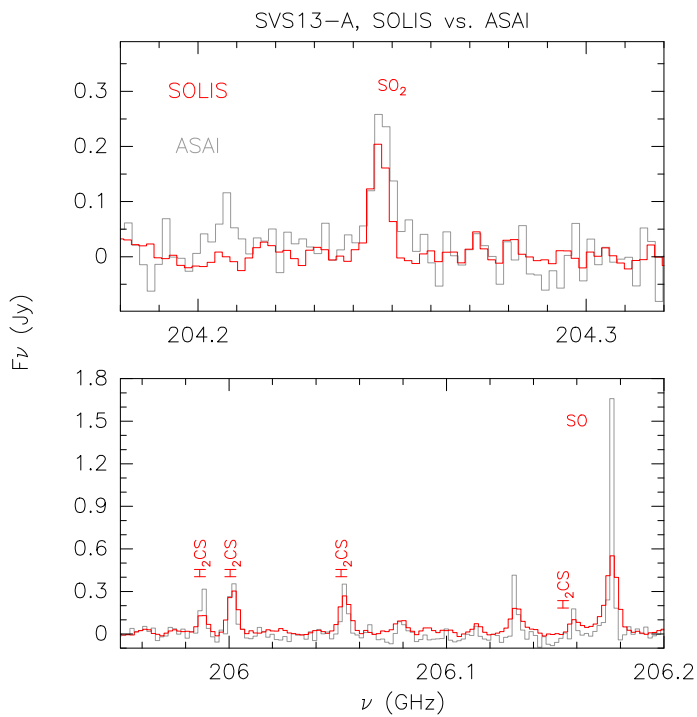}}
\caption{Zoom-in of Fig. A.2 to enlight selected emission lines of S-bearing species: comparison (in flux density scale) of the 1.4mm spectrum as observed using the IRAM 30-m antenna (grey; in the ASAI LP context \citet{Lefloch2018}) and that extracted from the present NOEMA SOLIS maps (red) from a circular region equal to the IRAM 30-m HPBW (12$\arcsec$). The ASAI spectrum has been smoothed to match the SOLIS velocity resolution (2 MHz, see Sect. 3). Red labels indicate the S-bearing species analysed in this paper (see Table 1).}
\end{figure}


\end{document}